\documentclass[12pt, draftclsnofoot, onecolumn]{IEEEtran}
\usepackage{amssymb}
\usepackage{cite}
\usepackage{graphicx}
\usepackage{array}
\usepackage{multirow}
\usepackage{amsmath}
\usepackage{stfloats}
\usepackage{graphicx}
\usepackage{subfigure}
\usepackage{tabularx}
\usepackage{epsfig,epsf,balance,cite}
\usepackage[caption=false,font=normalsize,labelfont=sf,textfont=sf]{subfig}
\usepackage{setspace}
\usepackage{bm}
\usepackage{textcomp}
\usepackage{algorithmic}
\usepackage{algorithm}
\usepackage{subfigure}
\usepackage{caption}
\usepackage{graphicx}
\usepackage{setspace}
\usepackage{url}
\usepackage{verbatim}
\usepackage{graphicx}
\usepackage{cite}
\usepackage{amssymb}
\usepackage{color}
\usepackage{xpatch}
\definecolor{shadecolor}{RGB}{0,0,255}
\definecolor{blue}{RGB}{0,0,255}
\newtheorem{theorem}{Theorem}

\newtheorem{lemma}{Lemma}

\makeatletter

\ExplSyntaxOn
\cs_new:Npn \bibColoredItems #1#2
{
	\clist_map_inline:nn {#2} { \cs_new:cpn {bib@colored@##1} {#1} } 
}
\ExplSyntaxOff

\newcommand\bib@setcolor[1]{%
	\ifcsname bib@colored@#1\endcsname
	\expanded{\noexpand\color{\csname bib@colored@#1\endcsname}}%
	\else
	\normalcolor
	\fi
}

\xpatchcmd\@bibitem
{\item}
{\bib@setcolor{#1}\item}
{}{\fail}

\xpatchcmd\@lbibitem
{\item}
{\bib@setcolor{#2}\item}
{}{\fail}
\makeatother

\begin{document}
	
\title{Two-Timescale Design for Reconfigurable Intelligent Surface-Aided URLLC}
	
\author{Qihao Peng, Hong Ren, Cunhua Pan,~\IEEEmembership{Senior Member,~IEEE}, \\
	 Maged Elkashlan,~\IEEEmembership{Senior Member,~IEEE}, Ana Garc{\' i}a Armada,~\IEEEmembership{Fellow,~IEEE}, \\
	 and Petar Popovski,~\IEEEmembership{Fellow,~IEEE}.
		\thanks{Q. Peng and M. Elkashlan are with the School of Electronic Engineering and Computer Science at Queen Mary University of London, U.K. (e-mail: \{q.peng, maged.elkashlan\}@qmul.ac.uk). H. Ren and C. Pan are with National Mobile Communications Research Laboratory, Southeast University, Nanjing, China. (e-mail:\{hren,cpan\}@seu.edu.cn). Ana Garc{\' i}a Armada is with Department of Signal Theory and Communications, University Carlos III of Madrid, Avda. de la Universidad 30 (28911 Legan{\' e}s) Spain (email: agarcia@tsc.uc3m.es). Petar Popovski is with the Department of Electronic Systems, Aalborg University, 9220 Aalborg, Denmark (e-mail: petarp@es.aau.dk). (Corresponding author: Hong Ren)} 
	}
	
	\maketitle
	
\begin{abstract}
	In this paper, to tackle the blockage issue in massive multiple-input-multiple-output (mMIMO) systems, a reconfigurable intelligent surface (RIS) is seamlessly deployed to support devices with ultra-reliable and low-latency communications (URLLC). The transmission power of the base station and the phase shifts of the RIS are jointly devised to maximize the weighted sum rate while considering the spatially correlation and channel estimation errors. Firstly, \textcolor{black}{the relationship between the channel estimation error and spatially correlated RIS's elements is revealed by using the linear minimum mean square error}. Secondly, based on the maximum-ratio transmission precoding, a tight lower bound of the rate under short packet transmission is derived. Finally, the NP-hard problem is decomposed into two optimization problems, where the transmission power is obtained by geometric programming and phase shifts are designed by using gradient ascent method. Besides, we have rigorously proved that the proposed algorithm can rapidly converge to a sub-optimal solution with low complexity. Simulation results confirm the tightness between the analytic results and Monte Carlo simulations. Furthermore, the two-timescale scheme provides a practical solution for the short packet transmission.
\end{abstract}	
	

\section{Introduction}
Industry 4.0 advocates a paradigm of smart factories where interconnected industrial Internet of Things (IIoT) enable the realization of product manufacturing and various tasks \cite{li2017review}. Through industrial control networks (ICN), various components, such as sensors and robots, can collaborate efficiently with the aid of real-time and precise commands. Conventionally, the wired network infrastructure (i.e., copper or fiber) is adopted to enable real-time and precise control, which results in high expenditure on installation and maintenance. In contrast to wired network infrastructure, wireless networks are not only less costly but also more flexible for mobile IIoTs. However, a small amount of data commands need to be exchanged with decoding error probability (DEP) less than $10^{-6}$ and latency below 1 ms \cite{chang2019optimizing}, leading to challenges for wireless links to enable guaranteed connectivity performance.

Recently, the academic and industry communities have made significant contributions towards fulfilling the demanding requirements of \textcolor{black}{ultra-reliable and low-latency communications (URLLC)}. The achievable data rate under point-to-point communications was derived and the relationship between achievable data rate and blocklength was revealed in \cite{ref3}. Then, based on the approximated expression under finite blocklength, \textcolor{black}{the feasibility of short packet transmission has been analyzed in several communication systems}, such as orthogonal frequency-division multiplexing (OFDM) \cite{hamamreh2017ofdm}, massive multiple-input-multiple-output (mMIMO) \cite{ren2020joint,ostman2021urllc}, and cell-free mMIMO systems \cite{peng2022resource,nasir2021cell,shi2022decentralized,ICCC}. Despite the significant potential of current systems, such as mMIMO and OFDM, \textcolor{black}{in supporting low-latency and ultra-reliable services for multiple devices simultaneously}, ensuring device's quality of service remains a challenge because of the detrimental effects of severe blockages induced by tall buildings and mobile objects.

To tackle the above-mentioned hurdle, a potential solution is to integrate reconfigurable intelligent surfaces (RISs) with communication systems. This is due to the fact that RIS can tune a large number of passive scattering elements to enhance the useful signals or mitigate interfering signals at low cost \cite{wu2019intelligent,pan2021reconfigurable,zhi2022active,pan2022overview,wang2023ris}. The benefits of integrating RISs into communication systems have been investigated, such as unmanned aerial vehicle communications \cite{lu2021aerial,liu2020machine}, and cell-free systems \cite{al2021ris,ni2022partially}. However, the aforementioned contributions were primarily based on the Shannon Capacity, where the packet is assumed to be infinite.

Owing to the enhanced path gain with low energy consumption, \textcolor{black}{the RIS-aided short packet transmission have attracted extensive attention} \cite{ranjha2020urllc,hashemi2021average,chandra2022downlink,ren2021intelligent,hashemi2022joint}. Specifically, by designing the beamforming and optimizing blocklength, the authors of \cite{ranjha2020urllc} proposed a Nelder–Mead simplex to minimize the DEP by leveraging the benefits of RIS with UAV. By using the related Gamma distribution and approximation, the average rate under short packet transmission was analyzed in \cite{hashemi2021average}. Then, the authors of \cite{chandra2022downlink} further investigated the average rate under the correlated channel and electromagnetic interference. Ren \emph{et al.} of \cite{ren2021intelligent} performed a comprehensive analysis of the single user's average DEP and achievable data rate under various scenarios, such as the Rayleigh fading and the Rician fading channels. To satisfy multiple users' diverse requirements, the total blocklength was minimized in \cite{hashemi2022joint}.

Although previous studies have extensively investigated the advantages of RIS-enabled \textcolor{black}{finite blocklength transmission}, most of them focused on the instantaneous channel state information (CSI)-based beamforming. \textcolor{black}{This scheme, however, is impractical under the finite blocklength regime. Firstly, based on the results given in \cite{wang2020channel}, the blocklength used for channel estimation typically increases with the RIS's reflecting elements, which leads to prohibitively high pilot overhead. When the transmission blocklength is limited, the high pilot overhead means there are few blocklength left for data transmission, leading to severe data rate loss. As a result, it is more challenging to devise the optimal transmission schemes to strike a trade-off between channel estimation and data transmission. Secondly, the instantaneous CSI-based scheme demands that the base station (BS) determine the optimal resource allocation based on the instantaneous CSI. Therefore, information feedback and power control need to be executed frequently before each data transmission, which leads to high signal processing latency and makes it challenging to satisfy the low-latency requirements.}

To address these practical challenges, a promising solution named the two-timescale scheme was proposed in \cite{han2019large}, and was further analyzed in \cite{kammoun2020asymptotic,hu2021two,zhi2022two}. \textcolor{black}{Rather than relying on individual channels, the two-timescale scheme devises phase shifts using the statistical CSI of the BS-RIS-user channel. Such an approach facilitates channel estimation akin to the RIS-free systems, thereby considerably reducing the pilot overhead. Moreover, the phase design of the RIS can be determined relying on the large-scale fading factors, which eliminates the requirement for information feedback and resource allocation operations until the long-term CSI has changed. This approach significantly reduces information exchange and processing latency.} However, the above-mentioned contributions of \cite{kammoun2020asymptotic,han2019large,hu2021two,zhi2022two} focused on the infinite blocklength, and the short packet transmission scheme under the two-timescale scheme remains an open problem.

Against the above background, we integrate the RIS with an mMIMO system and devise the downlink two-timescale scheme while accounting for both channel estimation errors and the spatial correlation among the reflecting elements. First, we reveal the relationship between the normalized mean square error (NMSE) and spatially correlated RIS's elements by the linear minimum mean square error (LMMSE). Then, a tight lower bound for the ergodic data rate is derived in closed form by relying on statistical CSI, which is more applicable for mMIMO systems with large-scale antennas. Finally, an effective algorithm is proposed to maximize the system performance under the short packet transmission. The specific contributions are given as follows:
\begin{enumerate}
	\item In view of the channel estimation errors and spatially correlated reflective elements, the two-timescale scheme is tailored for the RIS-enabled \textcolor{black}{finite blocklength transmission} under severe blockage scenarios, including indoor and urban communications. \textcolor{black}{The double Gaussian distributed channel is estimated, and the expression of NMSE reveals that, in comparison to the independent elements, the spatial correlation of RIS reflective elements contributes to improving the accuracy of estimated channels. Furthermore, a tight lower bound based on the maximum ratio transmission (MRT) precoding is derived, which offers an explicit closed-form expression for the power allocation and phase shifts design.} 
	\item  Based on the statistical CSI, weighted sum rate under short packet transmission is maximized while satisfying per-user requirements of data rate and DEP. To address this issue, \textcolor{black}{an effective alternating optimization is proposed, where the phase shifts and transmission power are optimized relying on the accelerated gradient ascent method and geometric programming (GP) \cite{ren2020joint,van2018joint}, respectively. Furthermore, the convergence of our algorithm is rigorously proved}, which is verified in the simulation results.
	\item  Extensive numerical results demonstrate the tightness between the analytic expression and ergodic rate obtained from Monte-Carlo simulations, and insightful analysis is drawn. The obtained results provide evidence that our proposed method achieves performance close to the ideal with remarkably low complexity. Furthermore, the two-timescale scheme demonstrated its superiority to that relying on the instantaneous CSI, which provides a practical solution for short packet transmission.
\end{enumerate}

The remaining sections are organized as follows. The system model is provided and a lower bound based on large-scale fading factors is presented in Section II. Then, the weighted sum rate under short packet transmission is maximized by devising the two-timescale scheme in Section III. In Section IV, extensive numerical results are presented. Finally, our conclusions are drawn in Section V. 

\section{System Model}
\subsection{Channel model}

\begin{figure}
	\centering
	\includegraphics[width=3.2in]{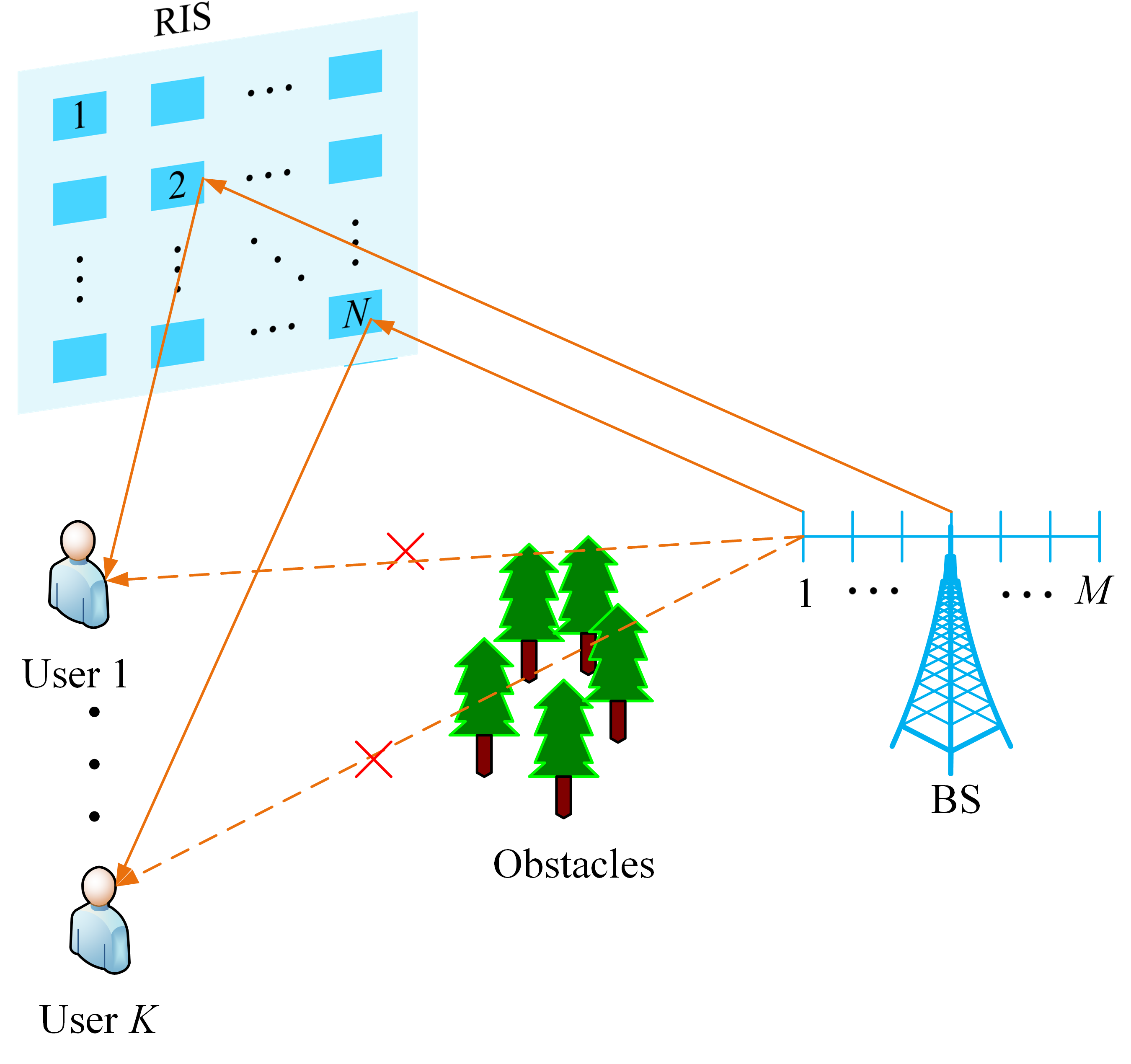}
	\caption{System model for a RIS-enabled mMIMO system.}
	\label{fig:system}
\end{figure}

In the system depicted in Fig. \ref{fig:system}, the $M$-antenna BS first estimates the uplink channel, and then performs beamforming techniques based on channel reciprocity to transmit signals to $K$ single-antenna devices. However, the direct links are susceptible to obstruction owing to numerous blocking obstacles (i.e., robots, trees, and vehicles). To address this blockage issue, an strategically deployed RIS is utilized to reflect the effective signal by tuning the $N$ passive and low-cost reflecting elements, thereby satisfying the diverse requirements of different devices.

The channel between the RIS and the $k$th device is denoted as ${\bf v}_k \in {\mathbb C}^{N \times 1}$, which is written as
\begin{equation}
\label{vk}
{{\bf{v}}_k} = \sqrt{ \beta _k^{{\rm{RU}}}}{\left( {{\bf{C}}_k^{{\rm{RU}}}} \right)^{\frac{1}{2}}}{{\bf{\tilde v}}_k},
\end{equation}
where $\beta _k^{{\rm{RU}}}$ is the distance-based path loss, ${{\bf{C}}_k^{{\rm{RU}}}} \in {\mathbb C}^{N \times N}$ (${{\bf{C}}_k^{{\rm{RU}}}} \succ {\bf 0}$) represents the RIS transmit correlation matrix, and ${{\bf{\tilde v}}_k} \sim {\cal CN}({\bf 0},{\bf I}_N)$ is the small-scale fading factors. Furthermore, the channel from the BS to the RIS ${\bf G} \in {\mathbb C}^{M \times N}$ is
\begin{equation}
\label{Gk}
{\bf{G}} = \sqrt{\beta^{\rm BR}}{\left( {{{\bf{C}}^{\rm{B}}}} \right)^{\frac{1}{2}}}{\bf{\tilde G}}{\left( {{{\bf{C}}^{\rm{R}}}} \right)^{\frac{1}{2}}},
\end{equation}
where $\beta^{\rm BR}$ is the large-scale fading factor, ${{\bf{C}}^{\rm{B}}} \in {{\mathbb C}^{M \times M}}$ (${{\bf{C}}^{{\rm{B}}}}  \succ {\bf 0}$) and ${{\bf{C}}^{\rm{R}}} \in {{\mathbb C}^{N \times N}}$ (${{\bf{C}}_k^{{\rm{R}}}}  \succ {\bf 0}$) are the BS transmit correlation matrix
and RIS receive correlation matrix for ${\bf G}$, respectively. As described in previous works \cite{hu2021two,abrardo2021intelligent,zhi2022two}, the channel ${\bf{\tilde G}} \in {{\mathbb C}^{M \times N}}$ follows the Rayleigh fading, and each element of ${\bf{\tilde G}}$ follows a complex Gaussian distribution ${\cal CN}(0,1)$. The cascaded channel from the BS to the $k$th device is expressed as ${{\bf{h}}_k} = {\bf{G\Phi }}{{\bf{v}}_k} \in {\mathbb C}^{M \times 1}$, where $ {\bf{\Phi }} = {\rm diag}\left\lbrace e^{j \theta_1},e^{j \theta_2},...,e^{j \theta_N}\right\rbrace$ is the phase shift matrix and $\theta_n \in \left[ {0,2\pi } \right)$ is the phase shift of the $n$th reflecting element. Then, the channel matrix is $ {\bf H} = [{\bf{h}}_1,{\bf{h}}_2,...,{\bf{h}}_K] \in {\mathbb C}^{M \times K}$.

\subsection{ \textcolor{black}{ Two-timescale and Channel Statistics}}
\begin{figure}
		\centering
		\includegraphics[width=3.2in]{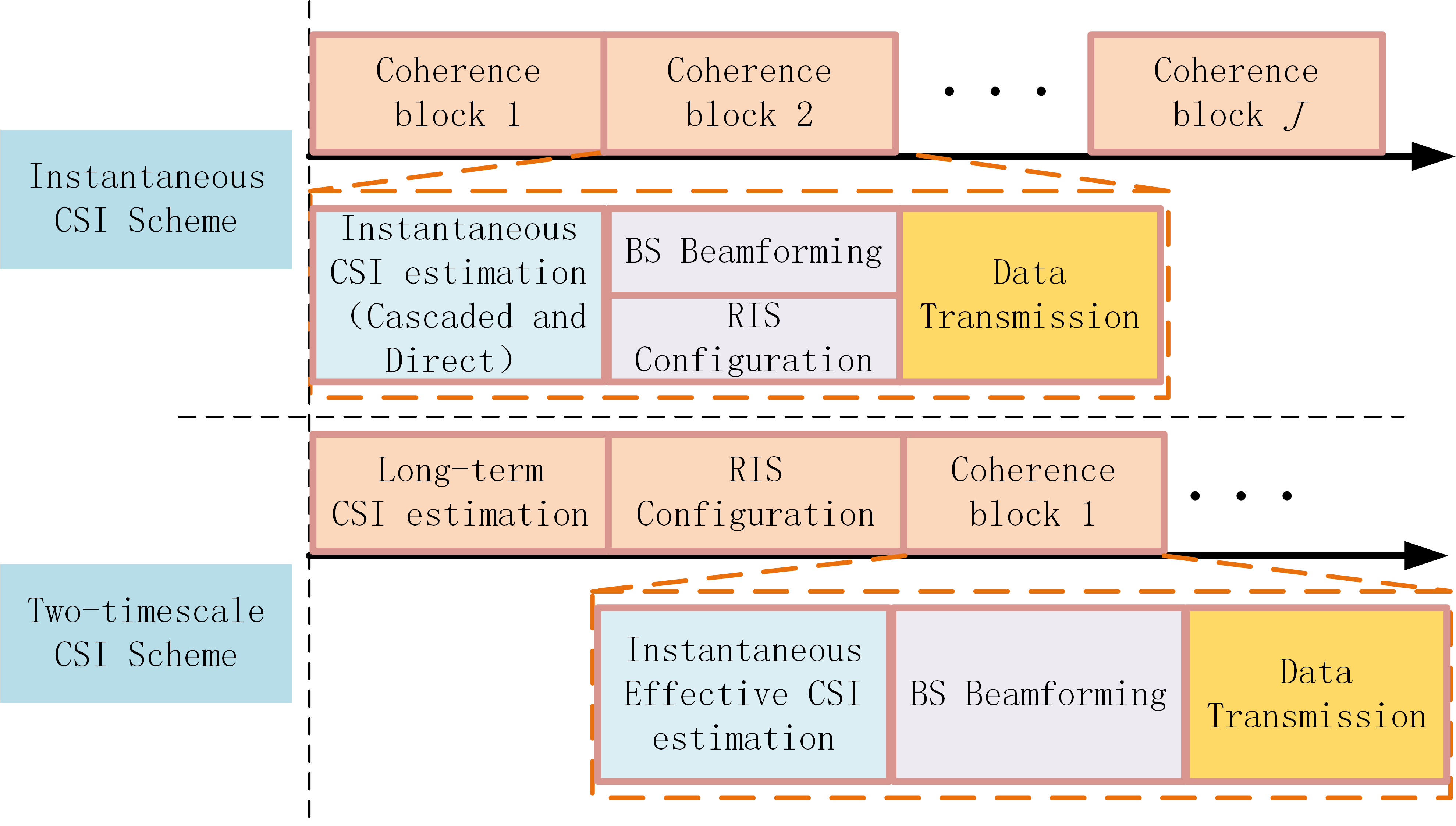}
		\caption{\textcolor{black}{Illustration of transmission designs based on Instantaneous CSI and long-term CSI ($J$ channel coherence blocks).}}
		\label{Twotimescale}
\end{figure}

\textcolor{black}{As can be seen from Fig. \ref{Twotimescale}, there are two transmission schemes. The first approach necessitates the separate estimation of BS-RIS and RIS-user channels, resulting in a notable pilot overhead for channel estimation. Besides, under short packet transmission, the channel blocklength $L$ is the product of bandwidth $B$ and transmission latency $T$, i.e., $L = B \times T$}. Once the bandwidth is determined, the channel blocklength is a linear function concerning the stringent transmission latency, leading to a finite blocklength. Therefore, it is challenging to devise the optimal transmission scheme based on the instantaneous CSI under short packet transmission.

In contrast, the two-timescale scheme can be implemented based on each user's effective channel in each coherence block, and thus the pilot overhead is equal to the number of users, which is the same as for mMIMO systems without RISs. Therefore, to satisfy the stringent latency, we investigate the two-timescale transmission scheme based on the channel statistics for the short packet transmission.

\textcolor{black}{To facilitate the transmission design, we need to obtain the channel statistics of the BS and the RIS. According to previous studies given in \cite{loyka2001channel,bjornson2020rayleigh}, the spatially correlation that is known as these channel statistics depends on the known correlation coefficient \cite{loyka2001channel} and the distance between the RIS's reflecting elements \cite{bjornson2020rayleigh}. Specifically, the BS transmit correlation matrix ${{\bf{C}}^{\rm{B}}}$ is based on the exponential correlation matrix \cite{loyka2001channel}. The $i$-th row and $j$-th column element of the BS's correlation matrix ${{\bf{C}}^{\rm{B}}}$ is given by:
	\begin{equation}
		c_{i,j}^B = \left\{ {\begin{array}{*{20}{c}}
				{{r^{j - i}},}&{i \le j}\\
				{{{\left( {c_{i,j}^B} \right)}^ * },}&{i > j}
		\end{array}} \right.,\left| r \right| \le 1
	\end{equation}
where $r$ is the (complex) correlation coefficient of neighboring receive branches. Furthermore, as stated in \cite{bjornson2020rayleigh}, the correlation $[{{\bf{C}}^{\rm{RU}}}]_{(i,j)}$ and $[{{\bf{C}}^{\rm{R}}}]_{(i,j)}$ between the $i$th element and $j$th element of the RIS is associated with the distance and wavelength $\lambda$, which is given by 
	\begin{equation}
		\label{correlated}
		c_{i,j} = {\rm sinc}\left( \frac{2|| {\bf u}_i - {\bf u}_j||}{\lambda})\right), 1\le i,j \le N,
	\end{equation}
\textcolor{black}{where $ {\rm sinc}(x) = \frac{\sin(\pi x)}{\pi x}$,${\bf u}_i = [0,{\rm mod}{(i-1,\sqrt{N})}d, \frac{\lfloor i-1\rfloor}{\sqrt{N}} d]^{T}$, $d$ is the width of each element.} $||{\bf u}_i - {\bf u}_j||$ denotes the distance from the $i$th to $j$th elements of the RIS. Therefore, these channel statistics can be determined by its manufacturing process and can be treated as known prior information.}

\subsection{Channel Estimation}
The uplink and downlink channels in time-division duplex (TDD) systems exhibit similar fading characteristics, known as channel reciprocity, and therefore the BS can utilize the results of uplink channel estimation to perform downlink precoding. For each transmission, the blocklength $L$ is limited with the product of bandwidth $B$ and transmission duration $T$. A segment of blocklength $\tau$ is used for estimating the uplink channel and the remaining portion for transmitting data. Furthermore, to prevent the $K$ devices from pilot contamination, it is essential to ensure orthogonality between the pilot sequences of each device. Consequently, the pilot length $\tau$ should be equal to or greater than $K$. Let us define ${\bf q}_k \in {\mathbb C}^{\tau \times 1}$ as the pilot sequence allocated for the $k$th device. Then, the expression of the received pilot signals ${\bf Y}_p \in M \times \tau$ at the BS is given by
\begin{equation}
	\label{pilot}
	{\bf Y}_p = \sqrt{\tau P_p} \sum\limits_{k = 1}^K {{{\bf{h}}_k} {\bf{q}}_k^H}  + {{\bf{N}}^p},
\end{equation}
where $P_p$ is the device's power for transmitting pilot. ${{\bf{N}}^p}$ is the additive noise, and each entry of ${{\bf{N}}^p}$ follows a complex normal distribution with zero mean and variance $\sigma^2$. The BS can obtain the cascaded channel to the $k$th device by using orthogonality of pilot sequence, which is given by 
\begin{equation}
	\label{hk}
	{{\bf y}_k^p} = {{\bf{h}}_k} + \frac{1}{{\sqrt {\tau P_p} }}{{\bf{N}}^p}{{\bf{q}}_k}.
\end{equation}

Given the receiving signal ${\bf y}_k^p$, the widely utilized estimation technique in RIS-free mMIMO systems is the minimum mean square error (MMSE) criteria. However, this method may be no longer optimal for the RIS-enabled MIMIO systems, since the cascaded channel $\bf H$ is double Gaussian distributed, rather than Gaussian distributed. As a result, in the following theorem, we estimate the aggregated channel by adopting the LMMSE method. This is due to the fact that the LMMSE method can obtain the sub-optimal estimation of receiving signal only relying on the statistical information, without knowledge of the exact channel distribution.

\begin{theorem}
	\label{theorem_estimate}
The estimated channel ${\bf{\hat h}}_k$ relying on LMMSE is given by
	\begin{equation}
		\label{hat_hk}
		\begin{split}
			{\bf {\hat h}}_k & =  {\beta _k^{{\rm{RU}}}\beta ^{{\rm{BR}}}{\rm{Tr}}\left\{ {\bf Z}_k \right\}{{\bf{C}}^{\rm{B}}}} \\
			& \times {\left( {\beta _k^{{\rm{RU}}}\beta^{{\rm{BR}}}{\rm{Tr}}\left\{ {\bf Z}_k \right\}{{\bf{C}}^{\rm{B}}} + \frac{\sigma^2}{{\tau P_p}}{{\bf{I}}_M}} \right)^{ - 1}}{\bf{y}}_k^p \\
			& \buildrel \Delta \over= {{\bf{R}}_k}{\bf{y}}_k^p,
		\end{split}
	\end{equation}
where ${\bf Z}_k = {{\bf{\Phi C}}_k^{{\rm{RU}}}{{\bf{\Phi }}^H}{{\bf{C}}^{\rm{R}}}}$. Then, the $k$th cascaded channel estimation error is 
\begin{equation}
	\label{NMSE}
	\begin{split}
		{\rm{NMSE}}_k 
		&= \frac{{{\rm{Tr}}\left\{ {\left( {{{\bf{I}}_M} - {{\bf{R}}_k}} \right){{\bf{C}}^{\rm{B}}}} \right\}}}{{{\rm{Tr}}\left\{ {{{\bf{C}}^{\rm{B}}}} \right\}}}.
	\end{split}
\end{equation}

\emph{Proof}: Please refer to Appendix \ref{Proof_estimation}. $\hfill\blacksquare$
\end{theorem}

With the given result in (\ref{hat_hk}), the estimated vector ${\bf {\hat h}}_k$ has the same dimension as the direct links in the RIS-free mMIMO systems. \textcolor{black}{Different from \cite{guo2022uplink}, we estimate the effective cascaded channel rather than individual channels, and thus the pilot overhead for estimating the aggregated channel does not increase with the number of the reflecting elements $N$, but instead the same as the number of devices $K$, which provides a more applicable solution for short packet transmission.}

\subsection{Achievable data rate under short packet transmission}
Shannon's coding theorem states that when the channel blocklength is infinite, the DEP has any impact on the rate. Conversely, if the blocklength is finite, then it would cause a considerable performance loss on the rate. Specifically, by treating the interference as a part of noise \footnote{\textcolor{black}{The channel hardening features of mMMIO result in SINR being heavily influenced by large-scale fading factors rather than small-scale ones. As a result, the instantaneous SINR can be considered equivalent to the expectation of SINR, such that we can approximate the equivalent channel as a Gaussian one, with SNR known at the transmitter, allowing the use of the finite blocklength formula for the Gaussian channels.}}, the data rate for small packet transmission is a function of blocklength, DEP, and SINR, which can be written as \cite{ref3}
\begin{equation}
	\label{urllc_rate}
	\setlength\abovedisplayskip{5pt}
	\setlength\belowdisplayskip{5pt}
	{R_k} \!\approx\! \left( {1 - \eta } \right){\log _2}\left( {1\! +\! \gamma_k} \right) \!-\! \frac{{{Q^{ - 1}}\left( {{\varepsilon _k}} \right)}}{{\ln 2}}\sqrt {\frac{{\left( {1 - \eta } \right){V_k\left(\gamma_k\right)}}}{L}} ,
\end{equation}
where $\eta = \tau / L$ means the ratio of pilot overhead, ${{\varepsilon _k}}$ denotes decoding error rate, ${\gamma}_k$ is the $k$th device's SINR,  $V_k = 1 - {\left( {1 + {\gamma_k}} \right)^{ - 2}}$ represents the channel dispersion in terms of the $k$th device's SINR ${\gamma}_k$. $Q\left( {{\varepsilon _k}} \right)$ is $\frac{1}{{\sqrt {2\pi } }}\int_{{\varepsilon _k}}^\infty  {{{\rm{e}}^{{{ - {t^2}} \mathord{\left/
					{\vphantom {{ - {t^2}} 2}} \right.
					\kern-\nulldelimiterspace} 2}}}{\rm{d}}t}$, and ${Q^{ - 1}}\left( {{\varepsilon _k}} \right)$ is the inverse function of $Q\left( {{\varepsilon _k}} \right)$.
				
Then, we aim to calculate the average data rate for short packet transmission, which is given by
\begin{equation}
	\setlength\abovedisplayskip{5pt}
	\setlength\belowdisplayskip{5pt}
	\label{rw_rate}
	\begin{split}
		{\bar R_k} & \approx \mathbb{E} \left\{ \frac{{1 - \eta }}{{\ln 2}}\left[ {\ln \left( {1 + {\gamma _k}} \right) - {\alpha _k} \sqrt {\frac{{\frac{2}{{{\gamma _k}}} + 1}}{{{{\left( {\frac{1}{{{\gamma _k}}} + 1} \right)}^2}}}} } \right] \right\}\\
		& \triangleq \frac{{1 - \eta }}{{\ln 2}} \mathbb{E} \left\{ f_k \left( \frac{1}{\gamma_k} \right)\right\},
	\end{split}
\end{equation}
where ${\alpha _k} = \frac{{{Q^{ - 1}}\left( {{\varepsilon _k}} \right)}}{{\sqrt {L\left( {1 - \eta } \right)} }}$, $f_k(x) = \ln(1+\frac{1}{x}) - \frac{{{Q^{ - 1}}\left( {{\varepsilon _k}} \right)}}{{\sqrt {L\left( {1 - \eta } \right)} }} \sqrt{\frac{2x + 1}{(1+x)^2}}$ is a function associated with the SINR and DEP of the $k$th device. Since the packet is delivered with extremely low latency, the distance-based path loss factors (i.e., large scale fading factors) remain constant with respect to the small-scale fading ones. Consequently, the expectation in (\ref{rw_rate}) is adopted to obtain the closed-form expression only relying on the large-scale fading factors. Such work, however, is a challenge, owing to the non-convex and non-concave expression under finite blocklength. To tackle this issue, previous studies mainly focused on approximating the rate through concave function approximation when dealing with short packet transmission \cite{ref20a,yu2021maximizing,abughalwa2022finite}. Similarly, we present a brief derivation for the lower bound of data rate.

Considering that the $k$th device's data rate always satisfies $R_k \ge 0$, we can equivalently obtain the following inequality, which is given by
\begin{equation}
	\setlength\abovedisplayskip{5pt}
	\setlength\belowdisplayskip{5pt}
	\label{a_region}
	{\alpha _k} \le \frac{{\left( {\frac{1}{{{\gamma _k}}} + 1} \right)\ln \left( {1 + {\gamma _k}} \right)}}{{\sqrt {\frac{2}{{{\gamma _k}}} + 1} }} \buildrel \Delta \over =  g\left( \frac{1}{{\gamma}_k} \right ),
\end{equation}
where $g(x) = \frac{(1+x)\ln(1+\frac{1}{x})}{\sqrt{2x + 1}}$. By checking the first-order derivative of $g\left( x \right )$, we can readily find that $g'(x) = \frac{x \ln(1 + \frac{1}{x}) - \frac{1}{x} - 2}{(2x + 1)^{\frac{3}{2}}}$ always satisfies $g'(x) < 0$, and thus $g\left( x \right )$ always decreases as $x$ increases. In addition, by combining (\ref{a_region}) with $f_k(x)$, we can readily obtain the feasible range of $f_{k} \left(x \right)$, represented as $0 \le x \le {g^{ - 1}}( {\alpha _k} )$. Based on these discussions, we prove that $f_{k} \left(x \right)$ is a decreasing and convex function when $x$ satisfies $x \in (0,g^{-1}({\alpha _k})] $. Then, by using Jensen's inequality, we can formulate the expression for the lower bound ${\hat R}_k$ of the $k$th device, which is given by
\begin{equation}
	\setlength\abovedisplayskip{5pt}
	\setlength\belowdisplayskip{5pt}
	\label{rate}
	{\bar R_k} \ge {\hat R}_k \triangleq \frac{{1 - \eta }}{{\ln 2}}{f_k}\left( {{1 \mathord{\left/
				{\vphantom {1 {{{\hat \gamma }_k}}}} \right.
				\kern-\nulldelimiterspace} {{{\hat \gamma }_k}}}} \right),
\end{equation}
where ${{\hat \gamma }_k}$ is ${{\hat \gamma }_k} = \frac{1}{\mathbb{E} \left( {{1 \mathord{\left/ {\vphantom {1 {{{\hat \gamma }_k}}}} \right. \kern-\nulldelimiterspace} {{{ \gamma }_k}}}} \right)}$. 

%\begin{lemma}
%	\label{x_region}
%	The function $f_k(x)$ exhibits properties of being both decreasing and convex when $x$ satisfies $x \in (0,g^{-1}({\alpha _k})] $ . %$\left \{ x | 0 < x \le {g^{ - 1}}\left( {\frac{{{Q^{ - 1}}\left( {{\varepsilon _k}} \right)}}{{\sqrt {L\left( {1 - \eta } \right)} }}} \right) \right\}$
%	
%	{\emph{Proof}}: Please refer to Appendix B in \cite{ref20a}. $\hfill\blacksquare$
%	
%\end{lemma}

In the following, a brief derivation for the lower bound using MRT precoding is presented. To eliminate the impact of small-scale fading parameters on the rate, we assume that the BS is equipped with large-scale antennas to achieve channel hardening and distance-based path loss is known in advance \cite{zhi2022two}. Based on the MRT precoding, the $k$th device's received signal is
\begin{equation}
	\label{received}
	\begin{split}
				r_k &= \mathbb{E}\underbrace {\left\{ {{\bf{h}}_k^H{{{\bf{\hat h}}}_k}\sqrt {p_k^d} } \right\}}_{{\rm{Desired \ signal}}}{x_k} \\
				& \quad- \underbrace {\left\{ {{\bf{h}}_k^H{{{\bf{\hat h}}}_k}\sqrt {p_k^d}  - \mathbb{E}\left\{ {{\bf{h}}_k^H{{{\bf{\hat h}}}_k}\sqrt {p_k^d} } \right\}} \right\}}_{{\rm{Signal \ leakage}}}{x_k} \\
		& \quad + \sum\limits_{k' \ne k}^K {\underbrace {{\bf{h}}_k^H{{{\bf{\hat h}}}_{k'}}\sqrt {p_{k'}^d} }_{{\rm{Multi-user\  interference}}}{x_{k'}}}  + n_k,
	\end{split}
\end{equation}
where $p_k^d$ is the transmission power, and $n_k$ is the additive noise with zero mean and variance $\sigma^2$. Then, the expression of the $k$th device's SINR is given by 
\begin{equation}
	\label{SINR}
	\gamma _k \! =\! \frac{p_k^d{{{\left| \mathbb{E} {\left\{ {{\bf{h}}_k^H{{{\bf{\hat h}}}_k}} \right\}} \right|}^2}}}{{ p_k^d{{\left| {{\bf{h}}_k^H{{{\bf{\hat h}}}_k} \!-\! \mathbb{E}\left\{ {{\bf{h}}_k^H{{{\bf{\hat h}}}_k} } \right\}} \right|^2}} \!+\! \sum\limits_{k' \ne k}^K {p_{k'}^d{{\left| {{\bf{h}}_k^H{{{\bf{\hat h}}}_{k'}} } \right|^2}}}  \!+\! \left| n_k\right|^2 }}.
\end{equation}

Since the mMIMO system has lots of transmission antennas, the law of large numbers is adopted to derive the lower bound of achievable data rate under short packet transmission.

\begin{theorem}
	\label{SINR_T}
	The lower bound for the $k$th device's averaged rate is
	\begin{equation}
		\setlength\abovedisplayskip{5pt}
		\setlength\belowdisplayskip{5pt}
		\label{SINR_LB}
		\begin{split}
		&{\hat R}_k  \buildrel \Delta \over= \frac{{1 - \eta }}{{\ln 2}}f_k \left( \frac{1}{{\hat \gamma }_k} \right), \\
		&\hat \gamma _k = \frac{p_k^d \left|{\rm Tr}\left\lbrace {\bf A}_k \right\rbrace  \right|^2 }{{\rm E}_{l,k} + \sum_{k' = 1}^{K}{p_{k'}^d {\rm UI}_{k,k'}} + \sigma^2},
		\end{split}
	\end{equation}
	where ${\bf A}_k$ is $ {\beta _k^{{\rm{RU}}}\beta^{{\rm{BR}}}{\rm{Tr}}\left\{ {\bf Z}_k \right\}{{\bf{R}}_k}{{\bf{C}}^{\rm{B}}}}$, ${\rm E}_{l,k}$ is $p_k^d{\left( {{\beta ^{\rm BR}}{\beta_k^{\rm RU}}} \right)^2}  {{\rm{Tr}}\left\{ {\bf Z}_k{\bf Z}_k \right\}{\rm{Tr}}\left\{ {{{\bf{C}}^{\rm{B}}}{\bf{R}}_k^H{{\bf{C}}^{\rm{B}}}{{\bf{R}}_k}} \right\}}$, and ${\rm UI}_{k,k'}$ is given by
	\begin{equation}
		\label{UIkk}
		\begin{split}
			{\rm UI}_{k,k'} &= \beta _{k'}^{\rm RU}\beta _k^{\rm RU}{\left( {{\beta ^{\rm BR}}} \right)^2} \Big( {{{\left| {{\rm{Tr}}\left\{ {{{\bf{C}}^{\rm{B}}}{{\bf{R}}_{k'}}} \right\}} \right|}^2}{\rm{Tr}}\left\{ {\bf Z}_k{\bf Z}_{k'} \right\}} \\
			&\quad+ {{\rm{Tr}}\left\{ {\bf Z}_{k'} \right\}{\rm{Tr}}\left\{ {{{\bf{C}}^{\rm{B}}}{{\bf{R}}_{k'}}{{\bf{C}}^{\rm{B}}}{\bf{R}}_{k'}^H} \right\}{\rm{Tr}}\left\{ {\bf Z}_k \right\}}\Big) \\
			&\quad  + \frac{\sigma^2}{{\tau P_p}}{\beta ^{\rm BR}}\beta _k^{\rm RU}{\rm{Tr}}\left\{ {{{\bf{C}}^{\rm{B}}}{{\bf{R}}_{k'}}{\bf{R}}_{k'}^H} \right\}{\rm{Tr}}\left\{ {\bf Z}_k \right\}.
		\end{split} 
	\end{equation}
	
	{\emph{Proof}}: Please refer to Appendix \ref{Proof_LB}. $\hfill\blacksquare$
\end{theorem}

The expression in (\ref{SINR_LB}) confirms that the SINR depends on the already known large-scale fading factors, in contrast to the uncertain ergodic rate through Monte Carlo simulations. Furthermore, owing to large-scale antenna transmission arrays in mMIMO systems, the derived lower bound is closer to the ergodic rate. As a result, the BS can allocate the transmission power and devise phase shifts according to the large-scale fading factors, which significantly reduce the information feedback and processing latency.

\section{Phase Shift Design and Power Allocation} 
According to the analytical result given in (\ref{SINR_LB}), we aim to design a scheme of power allocation and phase shifts to maximize the system performance by using the long-term statistical CSI. 

\subsection{Problem formulation}
Since the weighted sum rate is one of the key performance indicators of wireless communication systems, our objective is to maximize the weighted sum rate of the RIS-aided mMIMO system while satisfying short packet transmission requirements of decoding error probability and latency.

First, the weight $w_k$ is assigned to the $k$th device to provide priority for the specified devices. Then, by designing the RIS's phase shifts and optimizing the power consumption of BS, we aim to maximize the weighted sum rate while satisfying the minimal per-user rate requirement. The problem is formulated as
\begin{subequations}
	\setlength\abovedisplayskip{5pt}
	\setlength\belowdisplayskip{5pt}
	\label{MRC_optimization}
	\begin{align}
		\mathop {\max }\limits_{\left\{ \theta_n \right\},\left\{ {p_{k}^d} \right\}} & \sum\limits_{k = 1}^K {{w_k}{{\hat R}_k} } \label{MRC_optimization_a}\\
		{\rm{s}}{\rm{.t}}{\rm{.}}\;\;\;\; & {{\hat R}_k } \ge R_k^{{\rm{req}}},\forall k,  \label{MRC_optimization_b}\\
		& \theta_n \in [0,2\pi),\forall n,  \label{MRC_optimization_c}\\
		& {\sum\limits_{K=1}^{K} p_{k}^d \le {P^d}}, \label{MRC_optimization_d}
	\end{align}
\end{subequations}
where $R_k^{{\rm{req}}}$ is the $k$th device's minimal requirement of rate and DEP, and ${P^d}$ denotes the total transmitting power. Specifically, the constraint (\ref{MRC_optimization_b}) implies that the $k$th device needs to satisfy two conditions simultaneously: it should receive a signal with the data rate not lower than $R_k^{\rm{req}}$, and its decoding error rate should not be larger than ${{\varepsilon _k}}$.

Since the weighted sum rate maximization is an NP-hard problem, finding the optimal solution requires an exhaustive search, which is computationally demanding and impractical for communication systems. Moreover, the presence of a non-convex expression of  data rate brings further complexity to solve Problem (\ref{MRC_optimization}). As a result, maximizing the weighted sum rate requires efficient algorithms, which, even though sub-optimal, perform well in practice. To address this issue, we present an alternating optimization approach that provides a sub-optimal solution to the problem with low complexity.

\subsection{Power Optimization}
Owing to the non-convex expression in (\ref{urllc_rate}), finding an exact solution to transmission power is computationally challenging even with the fixed RIS's phase shifts matrix $\bf \Phi$. To solve this problem, we apply an equivalent transformation to convert it into a more tractable form. Firstly, using the lower bound given in (\ref{rate}), we can readily transform the requirements on data rate and DEP into the required SINR, which is given by
\begin{equation}
	\setlength\abovedisplayskip{5pt}
	\setlength\belowdisplayskip{5pt}
	\label{required_SINR}
	\hat \gamma _k \ge \frac{1}{{f_k^{ - 1}\left( {\frac{{R_k^{{\rm{req}}}\ln 2}}{{1 - \eta }}} \right)}}.
\end{equation}
Then, we can equivalently reformulate Problem (\ref{MRC_optimization}) by introducing the slack variables ${\chi _k}$ and we have
\begin{subequations}
	\setlength\abovedisplayskip{5pt}
	\setlength\belowdisplayskip{5pt}
	\label{MRC_optimization_trans}
	\begin{align}
		\mathop {\max }\limits_{\left\{ {p_{k}^d} \right\},\left\{ {\chi_k} \right\}} & {\sum\limits_{k = 1}^K {{\bar w}_k\left[ {\ln \left( {1 + {\chi _k}} \right) - {\alpha _k} \sqrt{1 - (1+\chi_k)^{-2}}} \right]} } \label{MRC_optimization_trans_a}\\
		\text{s.t.} \;\;\;\;\; & \hat \gamma _k \ge {\chi _k},\forall k,  \label{MRC_optimization_trans_b}\\
		&{\chi _k} \ge \frac{1}{{f_k^{ - 1}\left( {\frac{{R_k^{{\rm{req}}}\ln 2}}{{(1 - \eta) }}} \right)}},\forall k, \label{MRC_optimization_trans_c} \\
		&{\rm{ \left(\ref{MRC_optimization_d}\right)}}, \label{MRC_optimization_trans_d}
	\end{align}
\end{subequations}
where ${\bar w}_k$ is ${w_k}\frac{{\left( {1 - \frac{\tau }{L}} \right)}}{{\ln 2}}$. \textcolor{black}{Finally, since each device's SINR can exceed $\frac{\sqrt{17}-3}{4}$ to meet the stringent requirements, we utilize Lemma 2 and Lemma 3 given in \cite{ref13d} to approximate the weighted lower bound of the data rate by an iterative approach}, which is given by
\begin{equation}
	\label{wk_appro}
	\begin{split}
	& {{\bar w}_k\left[ \ln(1+\chi_k) - {\alpha _k} \sqrt{1 - (1+\chi_k)^{-2}} \right]} \\
	\ge & {{\bar w}_k\left[ {{\rho ^{\left( i \right)}_{k}}\ln \left( {{\chi _k}} \right) + {\delta ^{\left( i \right)}_{k}} - {\alpha _k}{{\hat \rho }^{\left( i \right)}_{k}}\ln \left( {{\chi _k}} \right) - {\alpha _k}{{\hat \delta }^{\left( i \right)}_{k}}} \right]} \\
	\ge & {{\bar w}_k\left[ {\ln \left( {{\chi _k}} \right)^{({\rho ^{\left( i \right)}_{k}} - {\alpha _k}{{\hat \rho }^{\left( i \right)}_{k}})} + ({\delta ^{\left( i \right)}_{k}} - {\alpha _k}{{\hat \delta }^{\left( i \right)}_{k}}}) \right]},
	\end{split}
\end{equation}
where  ${\rho ^{\left( i \right)}_{k}}$ is $\frac{{\chi}_k^{\left( i \right)}}{{1 + {\chi}_k^{\left( i \right)}}}$, ${\delta ^{\left( i \right)}_{k}}$ is $\ln \left( {1 + {\chi}_k^{\left( i \right)}} \right) - {\rho ^{\left( i \right)}_{k}}\ln \left( {{\chi}_k^{\left( i \right)}} \right)$, ${{\hat \rho }^{\left( i \right)}_{k}}$ is $\frac{{{\chi}_k^{\left( i \right)}}}{{\sqrt {{{{\chi}_k^{\left( i \right)}}^2} + 2{\chi}_k^{\left( i \right)}} }} - \frac{{{\chi}_k^{\left( i \right)}\sqrt {{{{\chi}_k^{\left( i \right)}}^2} + 2{\chi}_k^{\left( i \right)}} }}{{{{\left( {1 + \chi_k^{\left(i\right)}} \right)}^2}}}$, and ${{\hat \delta }^{\left( i \right)}_{k}}$ is $ \sqrt {1 - \frac{1}{{{{\left( {1 +{\chi}_k^{\left( i \right)}} \right)}^2}}}}  - {{\hat \rho }^{\left( i \right)}_{k}} \ln \left( {{\chi}_k^{\left( i \right)}} \right)$ in the $i$th iteration. Besides, the equality holds only when $\chi_k = \chi_k^{\left(i \right)}$. Substituting the lower bound of objective function into Problem (\ref{MRC_optimization_trans}), the original problem is rewritten as
\begin{subequations}
	\setlength\abovedisplayskip{5pt}
	\setlength\belowdisplayskip{5pt}
	\label{MRC_optimization_trans1}
	\begin{align}
		\mathop {\max }\limits_{\left\{ {p_{k}^d} \right\},\left\{ {\chi_k} \right\}} & {\sum\limits_{k = 1}^K {{\bar w}_k\left[ {\ln \left( {{\chi _k}} \right)^{({\rho ^{\left( i \right)}_{k}} - {\alpha _k}{{\hat \rho }^{\left( i \right)}_{k}})} + ({\delta ^{\left( i \right)}_{k}} - {\alpha _k}{{\hat \delta }^{\left( i \right)}_{k}}}) \right]} } \label{MRC_optimization_trans_a1}\\
		\text{s.t.} \;\;\;\;\; &{\rm{ \left(\ref{MRC_optimization_trans_b}\right)}}, {\rm{ \left(\ref{MRC_optimization_trans_c}\right)}},{\rm{ \left(\ref{MRC_optimization_d}\right)}}. \label{MRC_optimization_trans_d1}
	\end{align}
\end{subequations}

By ignoring the constant terms $({\delta ^{\left( i \right)}_{k}} - {\alpha _k}{{\hat \delta }^{\left( i \right)}_{k}})$ in (\ref{MRC_optimization_trans_a1}), the maximum exponential summation can be equivalently transformed into a product maximization problem, denoted as $\prod\limits_{k = 1}^K {{\chi _k}^{{{\hat w}_k^{\left( i \right)}}}}$, where ${{{\hat w}_k^{\left( i \right)}}}$ is ${\bar w}_k\left( {{\rho ^{\left( i \right)}} - {\alpha _k}{{\hat \rho }^{\left( i \right)}}} \right)$. Based on these discussions, the sub-optimal problem in the $i$th iteration can be expressed as
\begin{subequations}
	\setlength\abovedisplayskip{5pt}
	\setlength\belowdisplayskip{5pt}
	\label{MRC_equal_formualtion}
	\begin{align}
		\mathop {\max }\limits_{\left\{ {p_{k}^d} \right\},\left\{ {{\chi _k}} \right\}} & \prod\limits_{k = 1}^K {{\chi _k}^{{{\hat w}_k^{\left( i \right)}}}}  \label{equal_formualtion_a}\\
		{\text{s.t.}}\;\;\;\;\;\;\;\; & {\rm{ \left(\ref{MRC_optimization_trans_b}\right)}}, {\rm{ \left(\ref{MRC_optimization_trans_c}\right)}},{\rm{ \left(\ref{MRC_optimization_d}\right)}}. \label{equal_formualtion_b}
	\end{align}
\end{subequations}
Based on these discussions, it is feasible to efficiently solve Problem (\ref{MRC_equal_formualtion}) using existing optimization tools like CVX.
\subsection{Phase Shift Optimization}
With fixed transmission power, the RIS can tune the phase shift to maximize the weighted sum rate, which can be written as
\begin{subequations}
	\setlength\abovedisplayskip{5pt}
	\setlength\belowdisplayskip{5pt}
	\label{MRC_optimization_fai}
	\begin{align}
		\mathop {\max }\limits_{\left\{ \theta_n \right\}} & \sum\limits_{k = 1}^K {{w_k}{{\hat R}_k} } \label{MRC_optimization_fai_a} \\
		\text{s.t.} \;\;\;\;\; & {\rm{ \left(\ref{MRC_optimization_b}\right)}}, {\rm{ \left(\ref{MRC_optimization_c}\right)}}. \label{MRC_optimization_fai_b}
	\end{align}
\end{subequations}
To solve Problem (\ref{MRC_optimization_fai}), we employ the gradient ascent method to find the solution to the RIS's phase shifts. For ease of expression, we define ${\bf \Phi}$ as ${\rm diag}({\bf b})$, where ${\bf b} = e^{j {\boldsymbol \theta}}$ and ${\boldsymbol \theta} = \left[ \theta_1, \theta_2,...,\theta_N\right] ^T$ is a real vector of phase shift. It is noteworthy to mention that our method is based on the real phase shifts ${\boldsymbol \theta}$ rather than the complex variable ${\bf b}$, which is different from the existing work \cite{kammoun2020asymptotic}. Specifically, the existing methods based on complex gradients may not satisfy the constraint (\ref{MRC_optimization_c}) and only provide the approximated solutions, which causes additional errors and performance loss. By contrast, our method directly obtains the real gradient of phase shifts without approximations, and thus there is no additional performance loss.

Then, the gradient of weighted sum rate in terms of ${\boldsymbol \theta}$ is derived by applying the chain rule, which is detailed as
\begin{equation}
	\label{fk_gradient}
	\begin{split}
		\frac{{\partial \sum\limits_{k = 1}^K {{w_k}{{\hat R}_k}} }}{{\partial \boldsymbol \theta }} & = \sum\limits_{k = 1}^K \Big( {{\bar w}_k}\frac{1}{{\left( {1 +{{\hat \gamma}_k}} \right)}}\frac{{\partial {{\hat \gamma}_k}}}{{\partial \boldsymbol \theta }} \\
		 &\quad- {{\tilde w}_k} \frac{1}{{{{\left( {1 + {{\hat \gamma}_k}} \right)}^3}\sqrt {1 - \frac{1}{{{{\left( {1 + {{\hat \gamma}_k}} \right)}^2}}}} }}\frac{{\partial {{\hat \gamma}_k}}}{{\partial \boldsymbol \theta}}\Big),
	\end{split}
\end{equation}
where ${{\tilde w}_k}$ is $ {\bar w}_k \alpha_k$. The $\frac{{\partial {{\hat \gamma}_k}}}{{\partial \boldsymbol \theta}}$ is given by
\begin{equation}
	\label{SINR_gradient}
	\begin{split}
		\frac{{\partial {{\hat \gamma}_k}}}{{\partial \boldsymbol \theta}} &=\frac{{\frac{{\partial {p_k^d \left|{\rm Tr}\left\lbrace {\bf A}_k \right\rbrace  \right|^2}}}{{\partial \boldsymbol \theta }}}}{{{{\rm E}_{l,k}} + \sum_{k' = 1}^{K}{p_{k'}^d {\rm UI}_{k,k'}} + \sigma^2}} \\
		&\quad- \frac{{{p_k^d \left|{\rm Tr}\left\lbrace {\bf A}_k \right\rbrace  \right|^2}\left( {\frac{{\partial {{\rm E}_{l,k}}}}{{\partial \boldsymbol \theta }} + \sum_{k' = 1}^{K}{ \frac{\partial p_{k'}^d {\rm UI}_{k,k'}}{\partial \boldsymbol \theta}} } \right)}}{{{{\left( {{{\rm E}_{l,k}} + \sum_{k' = 1}^{K}{p_{k'}^d {\rm UI}_{k,k'}}  + \sigma^2} \right)}^2}}},
	\end{split}
\end{equation}
Then, two useful lemmas are presented to calculate the gradient of terms.
\begin{lemma}
	\label{lemma2_gradient}
	For the given matrices ${\bf A}$ and ${\bf B}$, the gradient of ${\rm{Tr}}({\bf A \Phi B} {\bf \Phi}^H)$ relying on phase shifts can be expressed as
	\begin{equation}
		\label{lemma2}
		\begin{split}
			\frac{{\partial {\rm{Tr}}\left\{ {{\bf{A}} {\bf \Phi} {\bf{B}}{{\bf \Phi} ^H}} \right\}}}{{\partial \boldsymbol \theta }} &=  - j{{\bf \Phi} ^H}\left( {{\bf{A}} \odot {{\bf{B}}^T}} \right){\bf{b}} \\
			&\quad+ j{{\bf \Phi} ^T}\left( {{\bf{A}}^{T} \odot {{\bf{B}}}} \right){{\bf{b}}^ * } = {\bf u}_g({\bf A},{\bf B}).
		\end{split}		
	\end{equation}
	
	{\emph{Proof}}: Please refer to Appendix \ref{Proof_lemma2}. $\hfill\blacksquare$
\end{lemma}

\begin{lemma}
	\label{lemma3_gradient}
    For a given matrix $\bf X$, $\frac{{\rm{Tr}}\left\{ {\bf{X}}{\bf R}_k \right\}}{{\partial \boldsymbol{\theta}  }}$ is expressed as
    \begin{equation}
    	\label{lemma3}
    	\begin{split}
    		\frac{{\rm{Tr}}\left\{ {\bf{X}}{\bf R}_k \right\}}{{\partial \boldsymbol{\theta}  }} &= \beta _k^{{\rm{RU}}}\beta ^{{\rm{BR}}} \Big( {\rm{Tr}} \left\lbrace  {\bf X}{\bf W}_k  \right\rbrace \\
    		&\quad - \beta _k^{{\rm{RU}}}\beta ^{{\rm{BR}}}{\rm{Tr}}\left\{ {\bf Z}_k \right\} {\rm{Tr}} \left\lbrace  {\bf X}{\bf W}_k {\bf W}_k  \right\rbrace \Big) 
    		\\ &\quad \times {\bf u}_g({\bf C}^{\rm R},{\bf C}^{\rm RU}_k) = {\bf z}_g({\bf X},{\bf W}_k),
    	\end{split}
    \end{equation}
	where ${\bf W}_k$ is equal to ${{\bf{C}}^{\rm{B}}}{\left( {\beta _k^{{\rm{RU}}}\beta^{{\rm{BR}}}{\rm{Tr}}\left\{ {\bf Z}_k \right\}{{\bf{C}}^{\rm{B}}} + \frac{\sigma^2}{{\tau P_p}}{{\bf{I}}_M}} \right)^{ - 1}}$. 
	
	{\emph{Proof}}: Please refer to Appendix \ref{Proof_lemma3}. $\hfill\blacksquare$
\end{lemma}

In the following, we derive the gradients of other terms by using Lemma \ref{lemma2_gradient} and Lemma \ref{lemma3_gradient}.
\begin{lemma}
	\label{lemma4}
	Gradients of terms ${\left|{\rm Tr}\left\lbrace {\bf A}_k \right\rbrace  \right|^2}$, ${\rm{Tr}}\left\{ {\bf Z}_k{\bf Z}_k \right\}$, ${{\rm{Tr}}\left\{ {{{\bf{C}}^{\rm{B}}}{\bf{R}}_k^H{{\bf{C}}^{\rm{B}}}{{\bf{R}}_k}} \right\}}$, and ${\rm{Tr}}\left\{ {{{\bf{C}}^{\rm{B}}}{{\bf{R}}_{k'}}{\bf{R}}_{k'}^H} \right\}$ with respect to $\boldsymbol{\theta}$ are given by
	\begin{equation}
		\label{ak}
		\begin{split}
			\frac{{\partial {\left|{\rm Tr}\left\lbrace {\bf A}_k \right\rbrace  \right|^2}}}{{\partial \boldsymbol \theta }}
			 & = 2 \beta _k^{{\rm{RU}}}\beta^{{\rm{BR}}}{{\rm Tr}\left\lbrace {\bf A}_k\right\rbrace  }\Big({\rm{Tr}}\left\{ {\bf Z}_k \right\} {\bf z}_g({\bf C}^{\rm B},{\bf W}_k)   \\
			 &\quad + {\rm{Tr}}\left\{ {{\bf{R}}_k}{{\bf{C}}^{\rm{B}}} \right\} {\bf u}_g({\bf C}^{\rm R},{\bf C}^{\rm RU}_k) \Big) ,
		\end{split}
	\end{equation}
	\begin{equation}
		\label{zkzk}
		\begin{split}
			\frac{\partial {\rm{Tr}}\left\{ {\bf Z}_k{\bf Z}_k \right\}}{\partial \boldsymbol{\theta}} &= {{\bf{u}}_g}\left( {{{\bf{C}}^{\rm{R}}},{\bf{C}}_k^{{\rm{RU}}}{{\bf \Phi}^H}{{\bf{C}}^{\rm{R}}}{\bf \Phi} {\bf{C}}_k^{{\rm{RU}}}} \right) \\
			&\quad  + {{\bf{u}}_g}\left( {{{\bf{C}}^{\rm{R}}}{\bf \Phi} {\bf{C}}_k^{{\rm{RU}}}{{\bf \Phi} ^H}{{\bf{C}}^{\rm{R}}},{\bf{C}}_k^{{\rm{RU}}}} \right), 
		\end{split}
	\end{equation}
	\begin{equation}
		\label{crcr}
		\begin{split}
			\frac{\partial {{\rm{Tr}}\left\{ {{{\bf{C}}^{\rm{B}}}{\bf{R}}_k^H{{\bf{C}}^{\rm{B}}}{{\bf{R}}_k}} \right\}}}{\partial \boldsymbol{\theta}} = 2{{\bf{z}}_g}\left( {{{\bf{C}}^{\rm{B}}}{{\bf{R}}_k}{{\bf{C}}^{\rm{B}}},{{\bf{W}}_k}} \right), 
		\end{split}
	\end{equation}
and
	\begin{equation}
	\label{crr}
	\begin{split}
		\frac{\partial {\rm{Tr}}\left\{ {{{\bf{C}}^{\rm{B}}}{{\bf{R}}_{k'}}{\bf{R}}_{k'}^H} \right\}}{\partial \boldsymbol{\theta}} = 2{{\bf{z}}_g}\left( {{{\bf{C}}^{\rm{B}}}{{\bf{R}}_{k'}},{{\bf{W}}_{k'}}} \right). 
	\end{split}
	\end{equation}
	{\emph{Proof}}: The proof can be established by applying the chain rules and leveraging Lemma 2 and Lemma 3. Let us take ${{\rm Tr}\left\lbrace {\bf Z}_k {\bf Z}_k\right\rbrace }$ for an example:
	\begin{equation}
		\label{proof_crr}
		\begin{split}
			&\frac{{\partial {\rm{Tr}}\left\{ {{{\bf{C}}^{\rm{R}}}{\bf \Phi} {\bf{C}}_k^{{\rm{RU}}}{{\bf \Phi}^H}{{\bf{C}}^{\rm{R}}}{\bf \Phi} {\bf{C}}_k^{{\rm{RU}}}{{\bf \Phi} ^H}} \right\}}}{{\partial \boldsymbol \theta }} \\
			& = \frac{{\partial {\rm{Tr}}\left\{ {{\bf{A}}{\bf \Phi}{\bf{B}}{{\bf \Phi} ^H}} \right\}}}{{\partial \boldsymbol \theta}}\left| {_{{\bf{A}} = {{\bf{C}}^{\rm{R}}},{\bf{B}} = {\bf{C}}_k^{{\rm{RU}}}{{\bf \Phi} ^H}{{\bf{C}}^{\rm{R}}}{\bf \Phi} {\bf{C}}_k^{{\rm{RU}}}}} \right.\\
			& \quad + \frac{{\partial {\rm{Tr}}\left\{ {{\bf{A}}{\bf \Phi} {\bf{B}}{{\bf \Phi} ^H}} \right\}}}{{\partial \boldsymbol \theta }}\left| {_{{\bf{A}} = {{\bf{C}}^{\rm{R}}}{\bf \Phi} {\bf{C}}_k^{{\rm{RU}}}{{\bf \Phi} ^H}{{\bf{C}}^{\rm{R}}},{\bf{B}} = {\bf{C}}_k^{{\rm{RU}}}}} \right.\\
			&= {{\bf{u}}_g}\left( {{{\bf{C}}^{\rm{R}}},{\bf{C}}_k^{{\rm{RU}}}{{\bf \Phi}^H}{{\bf{C}}^{\rm{R}}}{\bf \Phi} {\bf{C}}_k^{{\rm{RU}}}} \right) \\
			&\quad+ {{\bf{u}}_g}\left( {{{\bf{C}}^{\rm{R}}}{\bf \Phi} {\bf{C}}_k^{{\rm{RU}}}{{\bf \Phi} ^H}{{\bf{C}}^{\rm{R}}},{\bf{C}}_k^{{\rm{RU}}}} \right).
		\end{split}
	\end{equation}

	 The other terms can be obtained by using a similar but cumbersome process, and therefore are omitted here for brevity. $\hfill\blacksquare$
\end{lemma}

With the given results, we can readily obtain the gradient vectors $\frac{{\partial {{\rm E}_{l,k}}}}{{\partial \boldsymbol \theta }}$ and $\frac{{\partial {{\rm UI}_{k,k'}}}}{{\partial \boldsymbol \theta }} $  by using Lemmas \ref{lemma2_gradient}, \ref{lemma3_gradient}, \ref{lemma4} and chain rule. In the following, we calculate the gradient vector $\frac{{\partial {{\rm E}_{l,k}}}}{{\partial \boldsymbol \theta }}$, which is given by
\begin{equation}
	\label{elk}
	\begin{split}
		\frac{{\partial {{\rm E}_{l,k}}}}{{\partial \boldsymbol \theta }} &= p_k^d{\left( {{\beta ^{\rm BR}}{\beta_k^{\rm RU}}} \right)^2}  {{\rm{Tr}}\left\{ {{{\bf{C}}^{\rm{B}}}{\bf{R}}_k^H{{\bf{C}}^{\rm{B}}}{{\bf{R}}_k}} \right\}}  \\
		 &\quad \times \Big( {{\bf{u}}_g}\left( {{{\bf{C}}^{\rm{R}}},{\bf{C}}_k^{{\rm{RU}}}{{\bf \Phi}^H}{{\bf{C}}^{\rm{R}}}{\bf \Phi} {\bf{C}}_k^{{\rm{RU}}}} \right)   \\
		  & \quad\quad + {{\bf{u}}_g}\left( {{{\bf{C}}^{\rm{R}}}{\bf \Phi} {\bf{C}}_k^{{\rm{RU}}}{{\bf \Phi} ^H}{{\bf{C}}^{\rm{R}}},{\bf{C}}_k^{{\rm{RU}}}} \right) \Big) \\ 
		 & \quad + 2p_k^d{\left( {{\beta ^{\rm BR}}{\beta_k^{\rm RU}}} \right)^2} {\rm{Tr}}\left\{ {\bf Z}_k{\bf Z}_k \right\} {{\bf{z}}_g}\left( {{{\bf{C}}^{\rm{B}}}{{\bf{R}}_k}{{\bf{C}}^{\rm{B}}},{{\bf{W}}_k}} \right).
	\end{split}
\end{equation}

Then, following a similar process as (\ref{elk}) and after some algebraic manipulations, the gradient vector of $\frac{{\partial {{\rm UI}_{k,k'}}}}{{\partial \boldsymbol \theta }} $ can be obtained, which is given by

\begin{equation}
	\label{ekk}
	\begin{split}
			\frac{{\partial {{\rm UI}_{k,k'}}}}{{\partial \boldsymbol \theta }}  &= 2\beta _{k'}^{\rm RU}\beta _k^{\rm RU}{\left( {{\beta ^{\rm BR}}} \right)^2} {\rm Tr}\left\{ {\bf Z}_k{\bf Z}_{k'} \right\} {\rm Tr}\left\{ {\bf C}^{\rm B} {\bf R}_{k'} \right\} {\bf z}_g({ {\bf C}^{\rm B}},{\bf W}_{k'}) \\
			&\quad  + \beta _{k'}^{\rm RU}\beta _k^{\rm RU}{\left( {{\beta ^{\rm BR}}} \right)^2} {{\left| {{\rm{Tr}}\left\{ {{{\bf{C}}^{\rm{B}}}{{\bf{R}}_{k'}}} \right\}} \right|}^2}  {{\bf{u}}_g}\left( {{{\bf{C}}^{\rm{R}}}{\bf \Phi} {\bf{C}}_k^{{\rm{RU}}}{{\bf \Phi} ^H}{{\bf{C}}^{\rm{R}}},{\bf{C}}_{k'}^{{\rm{RU}}}} \right) \\
			&\quad  + \beta _{k'}^{\rm RU}\beta _k^{\rm RU}{\left( {{\beta ^{\rm BR}}} \right)^2} {{\left| {{\rm{Tr}}\left\{ {{{\bf{C}}^{\rm{B}}}{{\bf{R}}_{k'}}} \right\}} \right|}^2} {{\bf{u}}_g}\left( {{{\bf{C}}^{\rm{R}}},{\bf{C}}_k^{{\rm{RU}}}{{\bf \Phi}^H}{{\bf{C}}^{\rm{R}}}{\bf \Phi} {\bf{C}}_{k'}^{{\rm{RU}}}} \right)\\
			&\quad  + \beta _{k'}^{\rm RU}\beta _k^{\rm RU}{\left( {{\beta ^{\rm BR}}} \right)^2} {{\rm{Tr}}\left\{ {{{\bf{C}}^{\rm{B}}}{{\bf{R}}_{k'}}{{\bf{C}}^{\rm{B}}}{\bf{R}}_{k'}^H} \right\}{\rm{Tr}}\left\{ {\bf Z}_k \right\}} {\bf u}_g({\bf C}^{\rm R},{\bf C}^{\rm RU}_{k'}) \\
			& \quad  + \beta _{k'}^{\rm RU}\beta _k^{\rm RU}{\left( {{\beta ^{\rm BR}}} \right)^2} {{\rm{Tr}}\left\{ {{{\bf{C}}^{\rm{B}}}{{\bf{R}}_{k'}}{{\bf{C}}^{\rm{B}}}{\bf{R}}_{k'}^H} \right\}{\rm{Tr}}\left\{ {\bf Z}_{k'} \right\}} {\bf u}_g({\bf C}^{\rm R},{\bf C}^{\rm RU}_{k}) \\
			& \quad  + 2\beta _{k'}^{\rm RU}\beta _k^{\rm RU}{\left( {{\beta ^{\rm BR}}} \right)^2} {{\rm{Tr}}\left\{ {\bf Z}_{k} \right\}{\rm{Tr}}\left\{ {\bf Z}_{k'} \right\}}  {{\bf{z}}_g}\left( {{{\bf{C}}^{\rm{B}}}{{\bf{R}}_{k'}}{{\bf{C}}^{\rm{B}}},{{\bf{W}}_{k'}}} \right) \\
			& \quad + \frac{\sigma^2}{{\tau P_p}}{\beta ^{\rm BR}}\beta _k^{\rm RU}{\rm{Tr}}\left\{ {{{\bf{C}}^{\rm{B}}}{{\bf{R}}_{k'}}{\bf{R}}_{k'}^H} \right\}{\bf u}_g({\bf C}^{\rm R},{\bf C}^{\rm RU}_{k}) \\
			& \quad + 2\frac{\sigma^2}{{\tau P_p}}{\beta ^{\rm BR}}\beta _k^{\rm RU}{\rm{Tr}}\left\{ {\bf Z}_k \right\}{{\bf{z}}_g}\left( {{{\bf{C}}^{\rm{B}}}{{\bf{R}}_{k'}},{{\bf{W}}_{k'}}} \right).
		\end{split}
\end{equation}

Finally, by substituting the results of (\ref{ak}), (\ref{elk}), and $\frac{{\partial {{\rm UI}_{k,k'}}}}{{\partial \boldsymbol \theta }}$ into (\ref{fk_gradient}), we obtain the weighted sum rate's gradient with respect to ${\boldsymbol \theta}$. We can continuously tune the phase shift to obtain the maximum system performance as the functions always increase along with the gradient's direction. \textcolor{black}{Furthermore, to reduce the number of iterations of the proposed algorithm, we adopt the accelerated gradient ascent algorithm, rather than the conventional gradient ascent one, and the detailed algorithm is presented in Algorithm \ref{MRC_algorithm}}.

Furthermore, to identify a feasible region for the NP-hard problem, the algorithm for initializing the phase shift design and transmission scheme is given by
\begin{subequations}
	\setlength\abovedisplayskip{5pt}
	\setlength\belowdisplayskip{5pt}
	\label{feasible}
	\begin{align}
		\mathop {\max }\limits_{\left\{ \theta_n \right\},\left\{ {p_{k}^d} \right\},\left\{ \chi_k \right\}, \varGamma } & \varGamma  \label{feasible_a}\\
		{\rm{s}}{\rm{.t}}{\rm{.}}\;\;\;\; & {\chi _k} \ge \frac{\varGamma}{{f_k^{ - 1}\left( {\frac{{R_k^{{\rm{req}}}\ln 2}}{{(1 - \eta) }}} \right)}},\forall k,  \label{feasible_b}\\
		& { \left| e^{j\theta_n}\right| = 1},\forall n,  \label{feasible_c}\\
		& {\sum\limits_{K=1}^{K} p_{k}^d \le {P^d}}. \label{feasible_d}
	\end{align}
\end{subequations}
Obviously, the feasible region can be determined only when $\varGamma \ge $ 1. The process of finding the feasible region follows the same alternating optimization algorithm described in Algorithm \ref{MRC_algorithm}, and therefore it is omitted for brevity.
\subsection{Algorithm Analysis}
\subsubsection{Convergence} To ensure that the algorithm converges to a locally sub-optimal solution, it is necessary to demonstrate that the objective function consistently increases with the number of iterations, denoted as ${\rm Obj}^{(i+1)} \geq {\rm Obj}^{(i)}$. ${\rm Obj}^{(i)}$ represents the objective function in the $i$th iteration. Given the phase shifts $\boldsymbol{\theta}^{(i)}$ in the $i$th iteration, the power allocation scheme is proved to approach the sub-optimal solution asymptotically. Defining the power vector ${\bf p}^{d,(i+1)}$ as the $(i+1)$th solution and the corresponding vector of SINRs as ${ \boldsymbol \chi}_k^{(i+1)}({\bf p}^{d,(i+1)},\boldsymbol{\theta}^{(i)})$, we have
\begin{equation}
	\label{power_ith}
	\begin{split}
		&\sum\limits_{k = 1}^K {\bar w}_k\Big[\ln \left( \chi_k^{(i+1)}({\bf p}^{d,(i+1)},\boldsymbol{\theta}^{(i)}) \right)^{({\rho ^{\left( i \right)}_{k}} - {\alpha _k}{{\hat \rho }^{\left( i \right)}_{k}})} \\
		& \quad + ({\delta ^{\left( i\right)}_{k}} - {\alpha _k}{{\hat \delta }^{\left( i \right)}_{k}}) \Big]  \\
		\ge & \sum\limits_{k = 1}^K {\bar w}_k\Big[\ln \left( \chi_k^{(i)}({\bf p}^{d,(i+1)},\boldsymbol{\theta}^{(i)}) \right)^{({\rho ^{\left( i \right)}_{k}} - {\alpha _k}{{\hat \rho }^{\left( i \right)}_{k}})} \\
		& \quad + ({\delta ^{\left( i\right)}_{k}} - {\alpha _k}{{\hat \delta }^{\left( i \right)}_{k}}) \Big]   \\
		=& \sum\limits_{k = 1}^K {{w_k}{{\hat R}_k({\bf p}^{d,(i)},\boldsymbol{\theta}^{(i)} })}.
	\end{split}
\end{equation}  

Then, we combine (\ref{wk_appro}) with (\ref{power_ith}) and prove that the objective function based on the $(i+1)$th power allocation is larger than that relying on the $i$th power allocation, which can be written as
\begin{equation}
	\label{power_iith}
	\begin{split}
		&\sum\limits_{k = 1}^K {{w_k}{{\hat R}_k({\bf p}^{d,(i+1)},\boldsymbol{\theta}^{(i)} })}\\
		 = &\sum\limits_{k = 1}^K {\bar w}_k\Big[\ln \left( \chi_k^{(i+1)}({\bf p}^{d,(i+1)},\boldsymbol{\theta}^{(i)}) \right)^{({\rho ^{\left( i+1 \right)}_{k}} - {\alpha _k}{{\hat \rho }^{\left( i+1 \right)}_{k}})} \\
		 & \quad + ({\delta ^{\left( i+1\right)}_{k}} - {\alpha _k}{{\hat \delta }^{\left( i+1 \right)}_{k}}) \Big]   \\
		 \ge & \sum\limits_{k = 1}^K {\bar w}_k\Big[\ln \left( \chi_k^{(i+1)}({\bf p}^{d,(i+1)},\boldsymbol{\theta}^{(i)}) \right)^{({\rho ^{\left( i \right)}_{k}} - {\alpha _k}{{\hat \rho }^{\left( i \right)}_{k}})} \\
		 & \quad + ({\delta ^{\left( i\right)}_{k}} - {\alpha _k}{{\hat \delta }^{\left( i \right)}_{k}}) \Big]   \\ 
		 =& \sum\limits_{k = 1}^K {{w_k}{{\hat R}_k({\bf p}^{d,(i)},\boldsymbol{\theta}^{(i)} })}.
	\end{split}
\end{equation} 

With the given power allocation ${\bf p}^{d,(i+1)}$, we know that the objective function increases along the direction of the gradient, and we have
\begin{equation}
	\label{phase_ith}
	\begin{split}
		\sum\limits_{k = 1}^K {{w_k}{{\hat R}_k({\bf p}^{d,(i+1)},\boldsymbol{\theta}^{(i+1)} })} \ge \sum\limits_{k = 1}^K {{w_k}{{\hat R}_k({\bf p}^{d,(i+1)},\boldsymbol{\theta}^{(i)} })}.
	\end{split}
\end{equation}

Finally, combining the results (\ref{phase_ith}) with (\ref{power_iith}), the proof of the convergence of the proposed algorithm is completed, which is detailed as
\begin{equation}
	\label{convengence}
	\begin{split}
		{\rm Obj}^{(i+1)} &= \sum\limits_{k = 1}^K {{w_k}{{\hat R}_k({\bf p}^{d,(i+1)},\boldsymbol{\theta}^{(i+1)} })} \\
		& \ge \sum\limits_{k = 1}^K {{w_k}{{\hat R}_k({\bf p}^{d,(i+1)},\boldsymbol{\theta}^{(i)} })} \\
		& \ge \sum\limits_{k = 1}^K {{w_k}{{\hat R}_k({\bf p}^{d,(i)},\boldsymbol{\theta}^{(i)} })} = {\rm Obj}^{(i)}.
	\end{split}
\end{equation}

\subsubsection{Algorithm Complexity}
The complexity of our proposed algorithm primarily relies on two factors: calculating the gradient and solving the geometric programming (GP) problem in each iteration. Specifically, the complexity of the gradient ascent is to calculate the first-order derivative of the objective function, denoted as $N_{\theta}$, while that of the GP problem depends on solving Problem (\ref{MRC_equal_formualtion}) during each iteration \cite{boyd2007tutorial}. Since GP problem involves $2K$ variables and $(2K + 1)$ constraints, the computational complexity can be approximated as ${\mathcal{O}}(N_{a} \times \max{(2K + 1)(2K)^{2}, N_{d}})$, where $N_{a}$ represents the number of alternating optimizations, and $N_{d}$ pertains to the complexity for obtaining the first-order and second-order derivatives in Problem (\ref{MRC_equal_formualtion}) \cite{van2018joint}.  Therefore, the complexity of our alternating optimization algorithm is ${\mathcal{O}} (N_{a} \times [\max\{(2K + 1)(2K)^{2}, N_{d}\} + N_{\theta}])$. 

\begin{algorithm}[t]
	\caption{Alternating Iterative Optimization Algorithm}
	\begin{algorithmic}[1]
		\label{MRC_algorithm}
		\STATE The iteration number $i$ and error tolerance $\zeta$ is initialized as 1 and 0.001, respectively;
		\STATE Solving Problem (\ref{feasible}) to initialize the transmission power ${\bf p}^{d,(i)} = [p^{d,(i)}_1,p^{d,(i)}_2,...,p^{d,(i)}_K]^T$ and phase shifts $\boldsymbol{\theta}^{(i)}$, calculate the weighted sum rate ${\rm{WSR}}^{\left(i\right)} = \sum\limits_{k = 1}^K {{w_k}{{\hat R}_k({\bf p}^{d,(i)},\boldsymbol{\theta}^{(i)} })}$. Set ${\rm{WSR}}^{\left(0\right)} = 0$;
		\WHILE {${{\left( {{\rm{WS}}{{\rm{R}}^{\left( i \right)}} - {\rm{WS}}{{\rm{R}}^{\left( {i - 1} \right)}}} \right)} \mathord{\left/
					{\vphantom {{\left( {{\rm{WS}}{{\rm{R}}^{\left( i \right)}} - {\rm{WS}}{{\rm{R}}^{\left( {i - 1} \right)}}} \right)} {{\rm{WS}}{{\rm{R}}^{\left( {i - 1} \right)}}}}} \right.
					\kern-\nulldelimiterspace} {{\rm{WS}}{{\rm{R}}^{\left( {i - 1} \right)}}}} \ge \zeta$}
		\STATE Update $\left \{ {{\hat w}^{\left( i \right)}_k},\forall k \right\}$;
		\STATE With given phase shifts $\boldsymbol{\theta}^{(i)}$, using the CVX package to solve Problem (\ref{MRC_equal_formualtion}) and obtain transmission power ${\bf p}^{d,(i+1)}$; 
		\STATE \textcolor{black}{Set $n = 0$ and $a_0 = 0$. Initialize the iteration number $\boldsymbol{\theta}^{(i)}_n$ randomly  and ${\bf t}_{-1} = \boldsymbol{\theta}^{(i)}_n$.}
		\WHILE {1} 
		\STATE \textcolor{black}{Calculate the gradient vector of $\boldsymbol{\theta}^{(i)}_n$, denoted as ${\bf{f'}} (\boldsymbol{\theta}^{(i)}_n) = \frac{{\partial \sum\limits_{k = 1}^K {{w_k}{{\hat R}_k({\bf p}^{d,(i+1)},\boldsymbol{\theta}) }} }}{{\partial \boldsymbol{\theta} }}\left| {_{\boldsymbol{\theta}  = {\boldsymbol{\theta}^{(i)}_n}}} \right.$;
		\STATE Using the backtracking line search obtain the step size $k_n$;
		\STATE ${\bf t}_{n} = \boldsymbol{\theta}^{(i)}_n + k_n {\bf{f'}} (\boldsymbol{\theta}^{(i)}_n)$, $a_{n+1} =  \frac{1 +\sqrt{4 a^2_{n} + 1}}{2}$;
		\STATE $\boldsymbol{\theta}^{(i)}_{n +1} = {\bf t}_{n} + \frac{(a_n - 1)({\bf t}_{n} - {\bf t}_{n-1})}{a_{i+1}}$ ;}
		\IF {$\sum\limits_{k = 1}^K {{w_k}{{\hat R}_k({\bf p}^{d,(i+1)},\boldsymbol{\theta}^{(i)}_{n+1}) }} - \sum\limits_{k = 1}^K {{w_k}{{\hat R}_k({\bf p}^{d,(i+1)},\boldsymbol{\theta}^{(i)}_{n}) }} < 1{\rm e}^{-5}$}
		\STATE $\boldsymbol{\theta}^{(i+1)} = \boldsymbol{\theta}^{(i)}_{n+1}$, break;
		\ENDIF
		\STATE Update $n = n + 1$;
		\ENDWHILE
		\STATE Update $i = i + 1$, obtain the weighted sum rate ${\rm{WSR}}^{\left(i\right)}$ based on given transmission power ${\bf p}^{d,(i)}$ and phase shifts $\boldsymbol{\theta}^{(i)}$;
		\ENDWHILE
	\end{algorithmic}
\end{algorithm}

\section{Simulation Results} 
This section presents numerical results that validate the accuracy of our rigorous derivations compared to Monte Carlo simulations and demonstrate the performance improvement achieved by our proposed two-timescale scheme. 
\subsection{Simulation Setup}
To tackle the blockage issue of cell-edge devices, a typical application scenario of the RIS-enabled mMIMO system is investigated. Specifically, an $M$-antenna BS is in central coordinates, and $K$ devices randomly distributed on a semicircle of radius $r_{d}=5$ m, with center at 10 meters from the RIS located in the two-dimensional plane [0,50]. The distance-based path-loss factors $\beta^{\rm BR}$ and $\beta_k^{\rm RU}$ are modeled as $\beta^{\rm BR} = \beta_{0}(d^{\rm BR})^{-2.2}$, $\beta^{\rm RU}_k = \beta_{0}(d^{\rm RU}_k)^{- 2.1}$, respectively, where $\beta_{0} = 10^{-2}$, $d^{\rm BR}$ (m) and $d^{\rm RU}_k$ (m) represent the distance from the BS to the RIS, and from the RIS to the $k$th device. These parameters are similar to those in \cite{wang2020channel}. The correlation matrix model in \cite{bjornson2020rayleigh} is employed. Furthermore, we assume that the communication is over the bandwidth of $B=2$ MHz and the packet should be delivered below latency $T = 0.1$ ms, and therefore the blocklength is $L = B \times T = 200$. The transmission power and pilot power are $P^d = 1$ W and $p^p = 0.1$ W, and the power of noise is $\sigma^2 = B \times 1.381 \times 10^{-23} \times 290 \times 10 ^{\frac{9}{10}}$ W. The weight assigned to the $k$th device $w_k$ is randomly generated over the interval (0, 1]. The required data rate and decoding error rate are $R_k^{{\rm{req}}} = 0.2$ bit/s/Hz and ${{\varepsilon _k}} = 10^{-7}$, $\forall k$.

\subsection{Channel Estimation Error}
\begin{figure}
	\centering
	\includegraphics[width=3.2in]{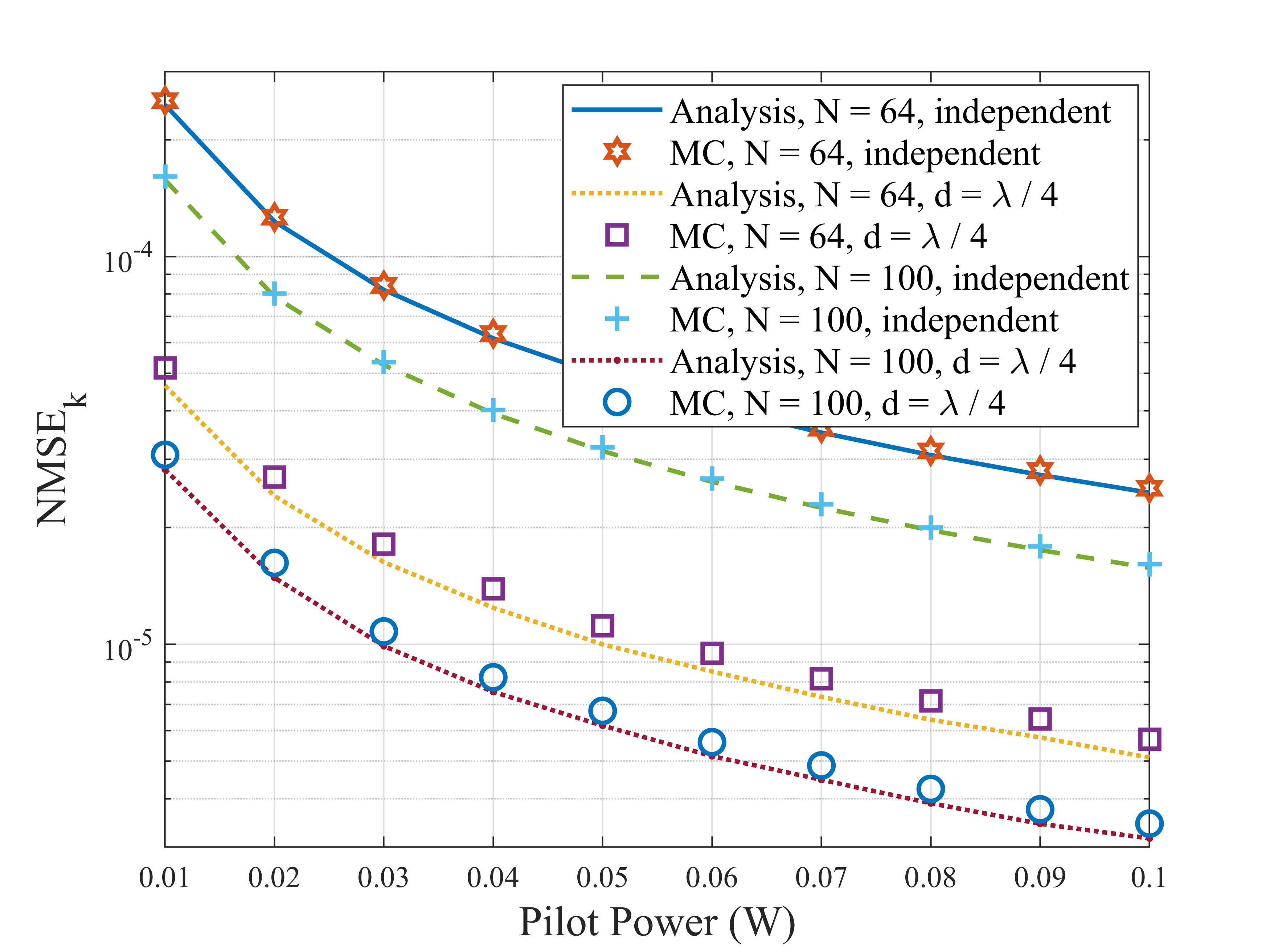}
	\vspace{-0.2cm}
	\caption{NMSE of device versus the pilot power with $M$ = 100.}
	\label{fig:NMSE}
\end{figure}
In Fig. \ref{fig:NMSE}, the NMSE based on LMMSE is evaluated in the cases of independent (i.e., ${\bf C}^{\rm B} = {\bf I}_M$ and ${\bf C}^{\rm R} = {\bf C}^{\rm RU}_k = {\bf I}_N$) and correlated channels, respectively. Apparently, increasing the pilot power can significantly improve the estimation performance. Besides, we observe that increasing the number of reflective elements of the RIS can effectively enhance the accuracy of the estimated channel. This phenomenon can be attributed to the fact that the increasing reflective elements $N$ can exploit the channel gain, which results in smaller channel estimation errors. More importantly, based on the results given in (6) and (7), both ${\bf R}_k$ and ${\bf C}^{\rm B}$ are non-diagonal matrices, and thus ${\rm Tr}\{{\bf R}_k{\bf C}^{\rm B}\}$ with correlated channel is larger than that of independent channel, which results in the improved estimation performance.

\subsection{Lower Bound and Ergodic Rate}
%\begin{figure*}[t]
%	\begin{minipage}[t]{0.49\linewidth}
%		\centering
%		\includegraphics[width=3.2in]{RIS_lower_bound.png}
%		\vspace{-0.2cm}
%		\caption{Weighted Sum Rate V.S. The Number of reflective elements of the RIS.}
%		\label{fig:MRC_LB}
%	\end{minipage}%
%	\hfill
%	\begin{minipage}[t]{0.49\linewidth}
%		\centering
%		\includegraphics[width=3.2in]{RIS_convergence.png}
%		\vspace{-0.2cm}
%		\caption{Weighted Sum Rate V.S. The Number of Iterations.}
%		\label{fig:FZF_LB}
%	\end{minipage}%
%\end{figure*}
\begin{figure}
\centering
\includegraphics[width=3.2in]{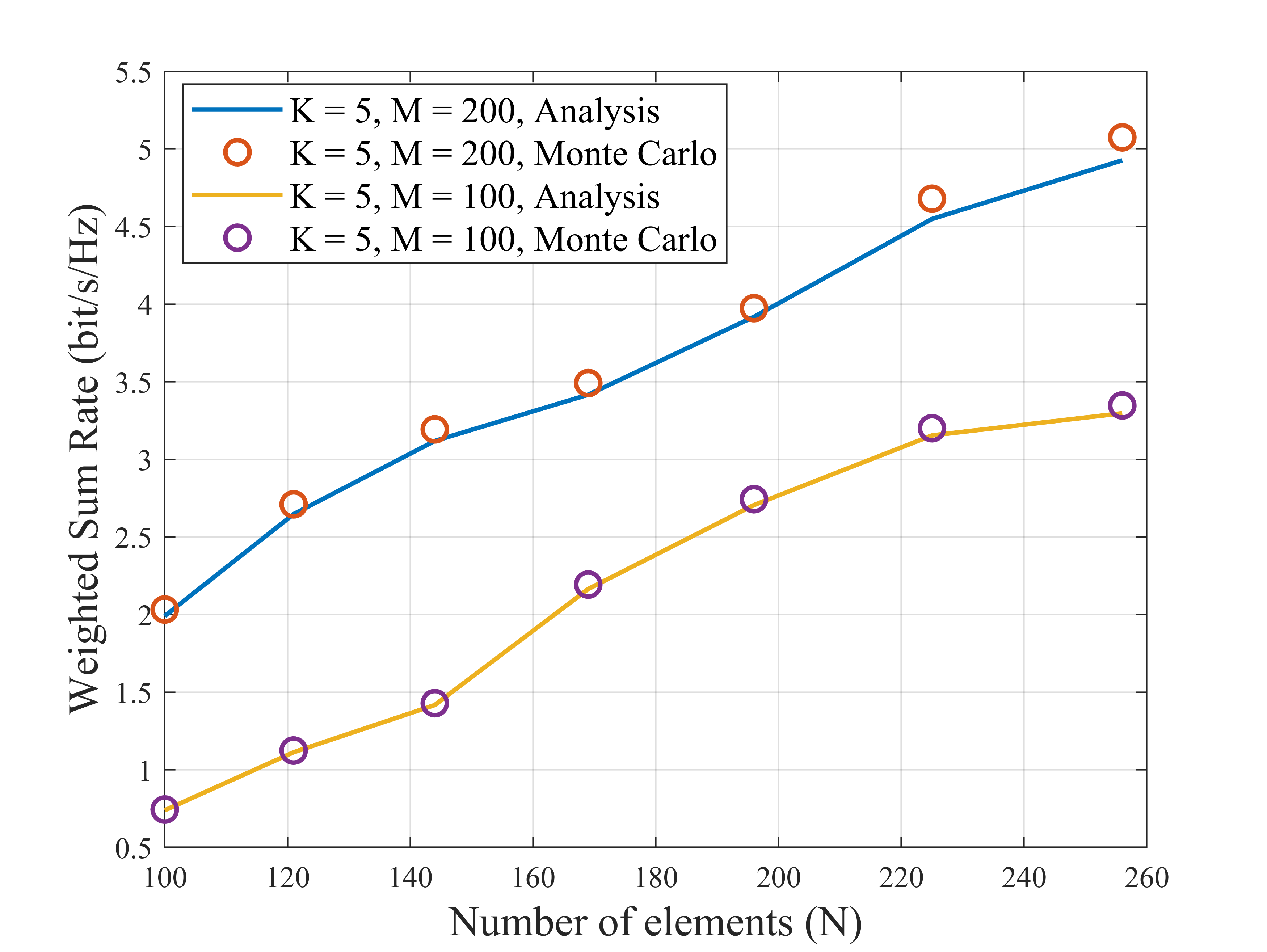}
\vspace{-0.2cm}
\caption{Weighted sum rate under versus number of reflective elements of RIS.}
\label{fig:MRC_LB}
\end{figure}

By averaging $10^4$ Monte Carlo simulation results, we investigate the tightness between our derived lower bound and achievable ergodic rate under the conditions of spatial distance $d = \frac{\lambda}{4}$ and transmission power $p_k^d = 0.2$ W, $\forall k$. As illustrated in Fig. \ref{fig:MRC_LB}, increasing the number of the antennas of BS and the number of reflecting elements of RIS can enhance the system performance. More importantly, our derived lower bound can approach the achievable ergodic rate in any case. As a result, compared to phase shift design and power optimization relying on the intricate formulation of the ergodic rate, the resource allocation relying on the derived closed-form expression is more practical and applicable.

%Furthermore, increasing more RIS elements has a significantly positive impact on boosting the weighted sum rate than installing more antennas to the BS, which demonstrates the benefits of integrating RIS into massive MIMO networks. For example, 100 antennas with 196 RIS elements can outperform 200 antennas with 100 RIS elements. 
\subsection{Convergence of the Alternating Optimization Algorithm}
%\begin{figure}
%\centering
%\includegraphics[width=3.2in]{RIS_convergence.png}
%\vspace{-0.2cm}
%\caption{Weighted sum rate versus the number of iterations.}
%\label{fig:FZF_LB}
%\end{figure}
\begin{figure}
	\centering
	\includegraphics[width=3.2in]{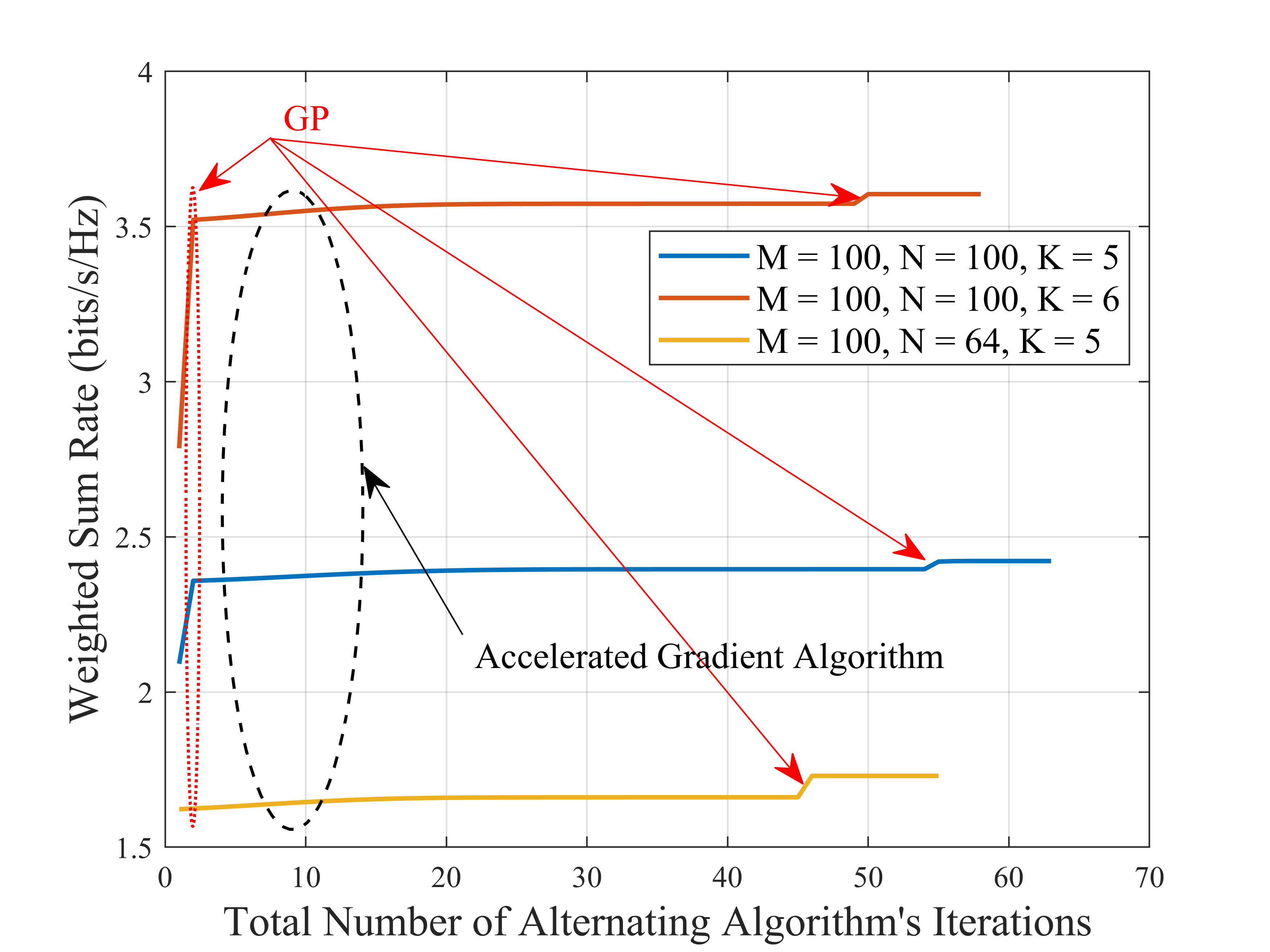}
	\vspace{-0.2cm}
	\caption{\textcolor{black}{Total Number of Alternating Algorithm's Iteration.}}
	\label{fig:FZF_LB}
\end{figure}
We depict the weighted sum rate of each iteration in Fig. \ref{fig:FZF_LB} to validate our rigorous convergence proof with spatial distance of $d = \frac{\lambda}{4}$, transmission power of $P^d = 1$ W, and pilot power of $p^p_k = 0.1$ W, $\forall k$. \textcolor{black}{It is observed from Fig. \ref{fig:FZF_LB} that the proposed algorithm converges within about 55 iterations (i.e., about 2 or 3 alternating optimizations), which confirms the accuracy of our rigorous derivations and demonstrates the effectiveness of our proposed algorithm.}

\subsection{Effect of Phase Shifts}
\begin{figure}
	\centering
	\includegraphics[width=3.2in]{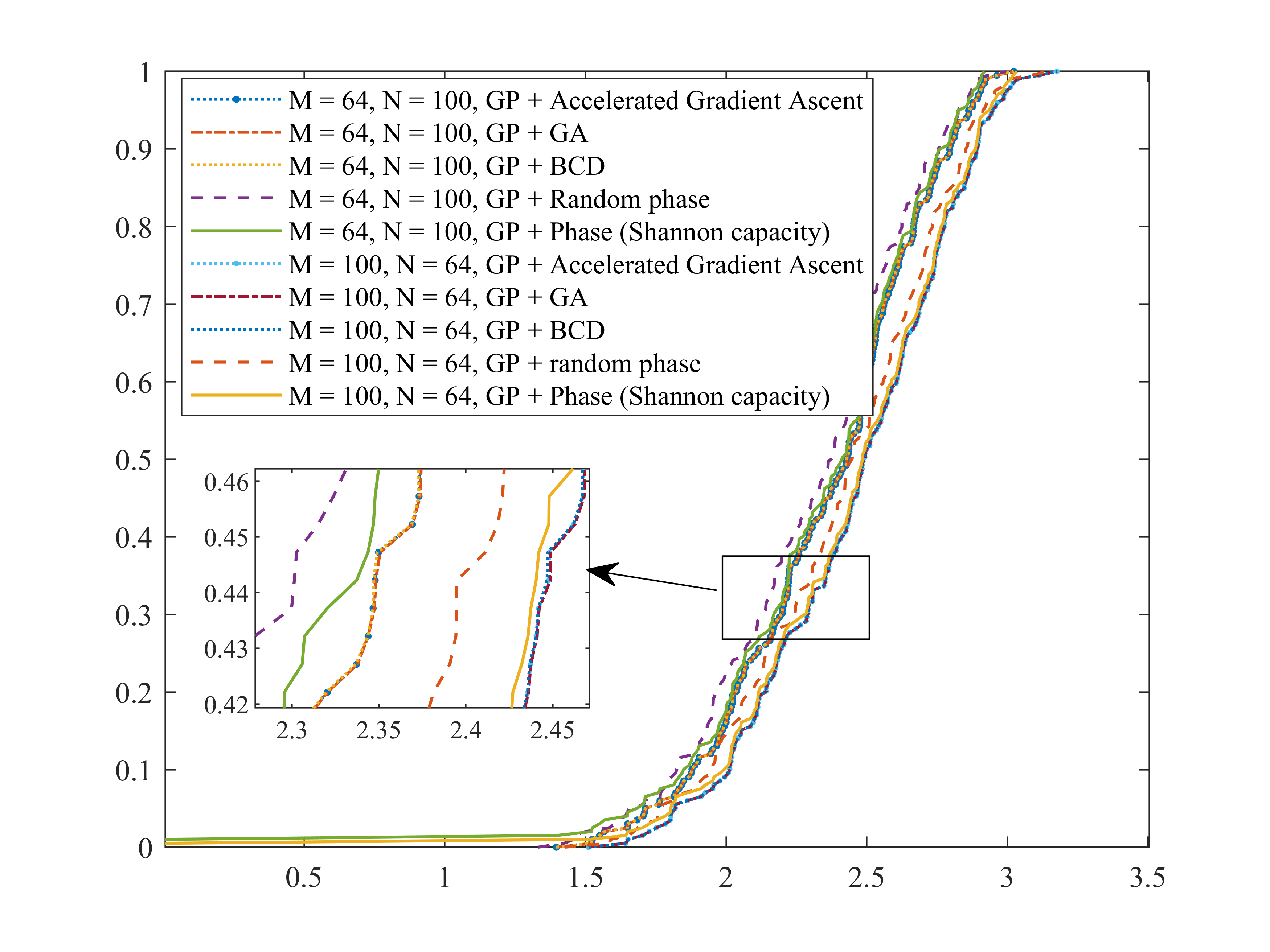}
	\vspace{-0.2cm}
	\caption{\textcolor{black}{CDF of weighted sum rate with $d = \frac{\lambda}{4}$ and $K = 5$.}}
	\label{fig:CDF}
\end{figure}
 
We evaluate the weighted sum rate's cumulative distribution function (CDF) to assess the system performance enhancement of our alternating optimization algorithm, as depicted in Fig. \ref{fig:CDF}. \textcolor{black}{In the scenario where no feasible region exists for our proposed algorithm, meaning that not all devices can fulfill the requirements of both the rate and DEP requirements, the corresponding weighted sum rates will be set to 0. Besides, we compare the proposed algorithm with the following algorithms:}
\begin{itemize}
	\item \textcolor{black}{GP + genetic algorithm (GA) \cite{zhi2022power}: GP is adopt to optimize the transmission power and GA is used for tuning phase shifts of the RIS.}
	\item \textcolor{black}{GP + block coordinate descent (BCD) \cite{wu2020joint}: GP and BCD are employed to devise the transmission power and phase shifts, respectively. }
	\item \textcolor{black}{GP + phase (Shannon capacity): The transmission power based on short packet transmission is obtained by using GP and the phase shifts are tuned relying on Shannon capacity.}
	\item \textcolor{black}{GP + random phase: We devise the transmission power and phase shifts by using GP and random phase, respectively.}
\end{itemize}
\textcolor{black}{Obviously, our proposed algorithm achieves superior performance over the scheme that only relies on power allocation and approaches the complex benchmark algorithms (i.e., combining GP with genetic GA and combining GP with BCD). Furthermore, The baseline based on Shannon capacity cannot guarantee all devises' requirement, as the phase shifts are obtained by Shannon capacity, ignoring the impact of decoding error probability. More importantly,} it is worth noting that, for the blocked devices, equipping more active antennas at the BS is more effective than equipping more passive reflective elements at the RIS.

\subsection{Effect of Number of Elements}
\begin{figure}
	\centering
	\includegraphics[width=3.2in]{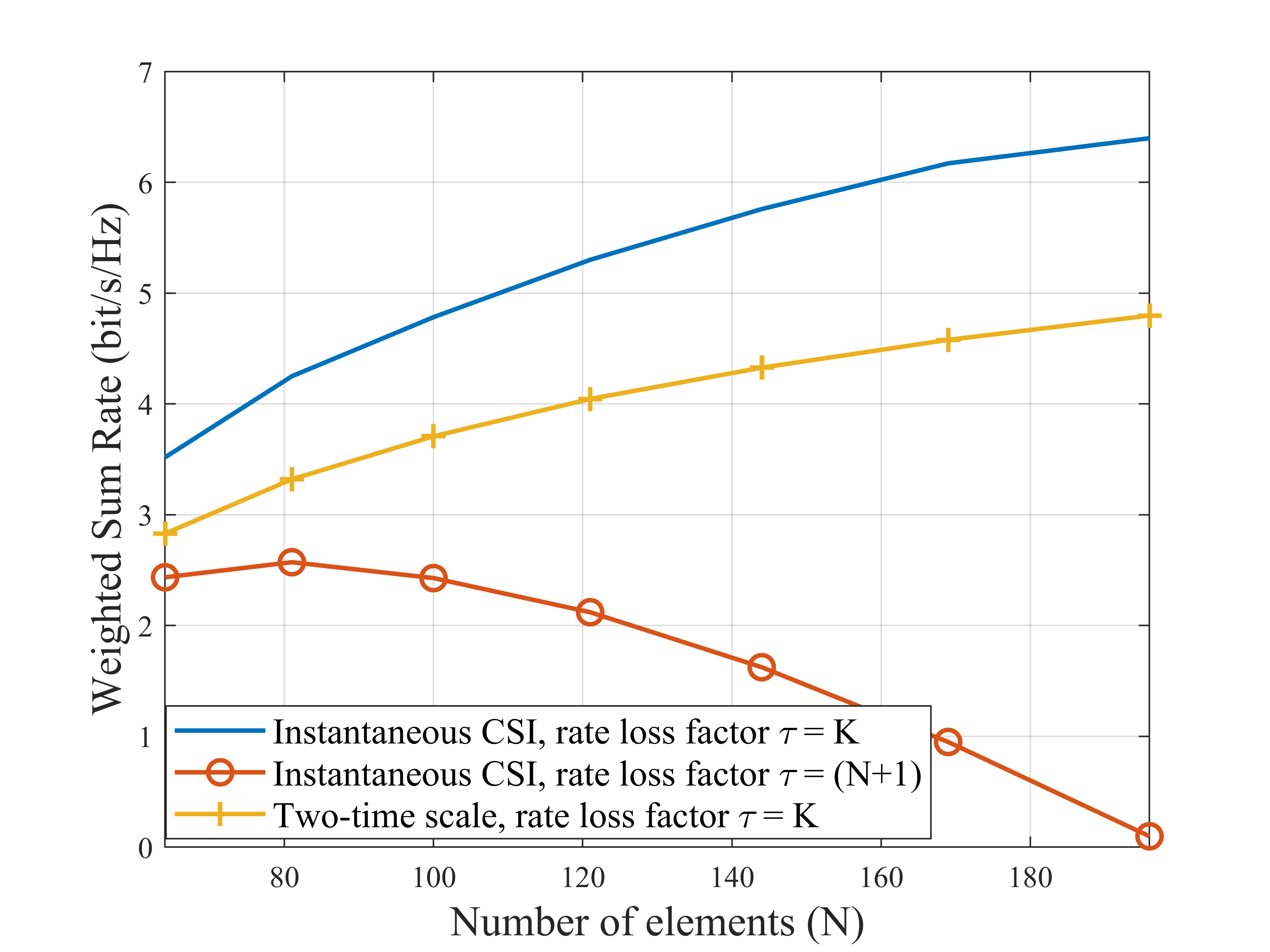}
	\vspace{-0.2cm}
	\caption{System performance versus various numbers of reflective elements with $d = \frac{\lambda}{4}$ and $K = 5$.}
	\label{fig:ins}
\end{figure}

In Fig. \ref{fig:ins}, to better demonstrate the superiority of the two-time scale scheme, we compare its performance with that relying on instantaneous CSI \textcolor{black}{scheme \cite{zhi2022two}, where the phase shift design can be obtained by GA.} Furthermore, \textcolor{black}{we also depict the ideal but unrealistic result where instantaneous CSI can be obtained by using $K$ blocklength and $N+1$ blocklength, respectively. For the ideal instantaneous scheme and the proposed two-timescale scheme (i.e., pilot overhead $\tau = K$), the weighted sum rates are both increasing with the increasing reflecting elements $N$, while the system performance relying on practical instantaneous CSI (i.e., $\tau = N+ 1$) improves slightly as the RIS's reflecting elements increase from 64 to 81 but degrades beyond 81.  Apparently, the ideal scheme based on instantaneous CSI exhibits superior performance, particularly for scenarios with large reflective elements $N$. This discrepancy arises because the instantaneous CSI-based scheme exploits optimal transmission design using both large-scale and small-scale fading factors in the $(L- K)$ blocklength. Furthermore, compared to the practical scheme upon instantaneous CSI (i.e., pilot overhead $\tau = N + 1$), the two-timescale scheme exhibits superior performance. This is because that when the number of reflective elements increases, the blocklength for channel estimation in the instantaneous CSI-based method grows as well.} As a result, with a large of passive reflective elements $N$, a significant portion of the blocklength is allocated for channel estimation, leaving only a limited number of symbols for data transmission. This results in a significant rate loss in the regime of large $N$. Therefore, Fig. \ref{fig:ins} demonstrates that the two-timescale scheme is a practical solution in the short packet transmission.

\section{Conclusion}
The two-timescale scheme for the RIS-enabled \textcolor{black}{finite blocklength transmission} was investigated while considering the channel estimation error and spatial correlation of reflective elements. Firstly, the channel was estimated relying on the LMMSE method, and the performance in terms of NMSE was analyzed. Then, a tight lower bound of data rate was derived for facilitating resource allocation. To tackle the intrinsically weighted sum rate maximization problem, GP and gradient descent methods were combined to devise the transmission scheme and phase shifts design. Simulation results validated that the BS can perform resource allocation relying on our derived results, which significantly reduces the information exchange and processing latency. Furthermore, compared to the practical case upon instantaneous CSI, the two-timescale scheme exhibits its superiority in terms of performance improvement and effectiveness.

Since the decoding error probability and latency are also the key performance indicators of wireless communication, it is valuable to investigate the transmission scheme and phase shifts to minimize the decoding error probability while satisfying the latency and data rate requirements, which would be left for our future work.
%\subsection{Effect of Blocklength}
%We investigate the effect of blocklength on the weighted sum rate, as illustrated in Fig. . 
%\bibColoredItems{black}{zhi2022power,wu2020joint,guo2022uplink,loyka2001channel}

\begin{appendices}

\section{Useful Lemmas}
\begin{lemma}
	Let ${\bf V} \in {\mathbb C}^{M \times N}$ be a complex matrix, where the $i$th row and $j$th column entry $[{\bf V}]_{(i,j)}$ follows a complex Gaussian distribution $\mathcal{CN}(0,1)$. For a given matrix ${\bf X} \in {\mathbb C}^{N \times N}$, the expectation of $ {\bf V} {\bf X} {\bf V}^H$ is expressed as
	\begin{equation}
		\label{lemma5}
		{\mathbb E}\left\lbrace {\bf V} {\bf X} {\bf V}^H \right\rbrace = {\rm Tr}\left\lbrace {\bf X}\right\rbrace {\bf I}_M.  
	\end{equation}

{\emph{Proof}}: The expectation of $[{\bf V} {\bf X} {\bf V}^H]_{i,j}$, $(i \ne j)$, can be given by
\begin{equation}
	\label{lemma51}
	\begin{split}
			&\mathbb{E}\left\{ {{{\left[ {{\bf{VX}}{{\bf{V}}^H}} \right]}_{i,j}}} \right\}\\
			 =& \mathbb{E}\left\{ {\sum\limits_{l = 1}^N {\sum\limits_{k = 1}^N {{{\left[ {\bf{V}} \right]}_{i,k}}{{\left[ {\bf{X}} \right]}_{k,l}}{{\left[ {{{\bf{V}}^H}} \right]}_{l,i}}} } } \right\} \\
		 =& \sum\limits_{l = 1}^N {\sum\limits_{k = 1}^N {\mathbb{E} \left\{ {{{\left[ {\bf{V}} \right]}_{i,k}}{{\left[ {{{\bf{V}}^H}} \right]}_{l,j}}} \right\}{{\left[ {\bf{X}} \right]}_{k,l}}} } = 0.
	\end{split}
\end{equation}

For $i = j$, we have
\begin{equation}
	\label{lemma52}
	\begin{split}
		&\mathbb{E}\left\{ {{{\left[ {{\bf{VX}}{{\bf{V}}^H}} \right]}_{i,i}}} \right\} \\
		=& \mathbb{E}\left\{ {\sum\limits_{l = 1}^N {\sum\limits_{k = 1}^N {{{\left[ {\bf{V}} \right]}_{i,k}}{{\left[ {\bf{X}} \right]}_{k,l}}{{\left[ {{{\bf{V}}^H}} \right]}_{l,i}}} } } \right\} \\
		 =& {\sum\limits_{k = 1}^N {\mathbb{E} \left\{ {{{\left[ {\bf{V}} \right]}_{i,k}}{{\left[ {{{\bf{V}}^H}} \right]}_{i,k}}} \right\}{{\left[ {\bf{X}} \right]}_{k,k}}} } = {\rm Tr} \left\lbrace \bf X \right\rbrace.
	\end{split}
\end{equation}

By combining the results of (\ref{lemma51}) and (\ref{lemma52}) into a matrix, we complete this proof. $\hfill\blacksquare$
\end{lemma}

\begin{lemma}
	Consider a matrix ${\bf V} \in {\mathbb C}^{M \times N}$, and each element of $[{\bf V}]$ is a complex Gaussian distribution $\mathcal{CN}(0,1)$. Given a matrix ${\bf X} \in {\mathbb C}^{N \times N}$ and a hermitian matrix ${\bf C} \in {\mathbb C}^{M \times M}$, we have
	\begin{equation}
		\label{lemma6}
		{\mathbb E}\left\lbrace {\bf V}^H {\bf C}{\bf V} {\bf X}{\bf V}^H{\bf C} {\bf V} \right\rbrace = {\rm Tr}\left\lbrace {\bf X}\right\rbrace{\rm Tr}\left\lbrace {{\bf C}^2}\right\rbrace {\bf I}_N + |{\rm Tr}\left\lbrace {\bf C}\right\rbrace|^2{\bf X}.  
	\end{equation}
	
	{\emph{Proof}}: Define ${\bf v}_i \in {\mathbb C}^{M \times 1}$ as the $i$th column of matrix $\bf V$. The expectation of $[{\bf V}^H {\bf C}{\bf V} {\bf X}{\bf V}^H{\bf C} {\bf V}]_{i,j} $, $(i \ne j)$, is
	\begin{equation}
		\label{lemma61}
		\begin{split}
			&\mathbb{E} \left\{ {{{\left[ {{{\bf{V}}^H}{\bf{CVX}}{{\bf{V}}^H}{\bf{CV}}} \right]}_{i,j}}} \right\} \\
			= &\mathbb{E} \left\{ {\sum\limits_{l = 1}^N {\sum\limits_{k = 1}^N {{\bf{v}}_i^H{\bf{C}}{{\bf{v}}_k}{{\left[ {\bf{X}} \right]}_{k,l}}{\bf{v}}_l^H{\bf{C}}{{\bf{v}}_j}} } } \right\} \\
%			& = \mathbb{E}\left\{ {{\bf{v}}_i^H{\bf{C}}{{\bf{v}}_i}{{\left[ {\bf{X}} \right]}_{i,j}}{\bf{v}}_j^H{\bf{C}}{{\bf{v}}_j}} \right\} \\ 
			=& {\left[ {\bf{X}} \right]_{i,j}}\mathbb{E}\left\{ {{\bf{v}}_i^H{\bf{C}}{{\bf{v}}_i}} \right\}\mathbb{E}\left\{ {{\bf{v}}_j^H{\bf{C}}{{\bf{v}}_j}} \right\} \\
			=& {\left[ {\bf{X}} \right]_{i,j}}{\left| {{\rm{Tr}}\left\{ {\bf{C}} \right\}} \right|^2}.
		\end{split}
	\end{equation}
	
	For $i = j$, we have
	\begin{equation}
		\label{lemma62}
		\begin{split}
			&\quad \mathbb{E}\left\{ {{{\left[ {{{\bf{V}}^H}{\bf{CVX}}{{\bf{V}}^H}{\bf{CV}}} \right]}_{i,i}}} \right\} \\
%			=& \mathbb{E} \left\{ {\sum\limits_{k = 1}^N {{\bf{v}}_i^H{\bf{C}}{{\bf{v}}_k}{{\left[ {\bf{X}} \right]}_{k,k}}{\bf{v}}_k^H{\bf{C}}{{\bf{v}}_i}} } \right\} \\
			& =  \mathbb{E}\left\lbrace {\bf{v}}_i^H{\bf{C}}{{\bf{v}}_i}{\left[ {\bf{X}} \right]_{i,i}}{\bf{v}}_i^H{\bf{C}}{{\bf{v}}_i}\right\rbrace  \\
			& \quad + \mathbb{E}\left\{ {\sum\limits_{k \ne i}^N {{\bf{v}}_i^H{\bf{C}}{{\bf{v}}_k}{{\left[ {\bf{X}} \right]}_{k,k}}{\bf{v}}_k^H{\bf{C}}{{\bf{v}}_i}} } \right\}\\
			 & \mathop  = \limits^{\left( a \right)}  {\left[ {\bf{X}} \right]_{i,i}}{\left| {{\rm{Tr}}\left\{ {\bf{C}} \right\}} \right|^2} + {\rm{Tr}}\left\lbrace \bf X \right\rbrace {\rm{Tr}}\left\lbrace {\bf C}^2 \right\rbrace,
		\end{split}
	\end{equation}
	where ${\left( a \right)}$ exploits $\mathbb{E} \left\{ {{{\left| {{\bf{v}}_i^H{\bf{C}}{{\bf{v}}_i}} \right|}^2}} \right\} = {\left| {{\rm{Tr}}\left\{ {\bf{C}} \right\}} \right|^2} + {\rm{Tr}}\left\{ {{{\bf{C}}^2}} \right\}$, the proof of which is detailed in Lemma 2 in \cite{bjornson2015massive}. 
	
	By combining the results of (\ref{lemma61}) and (\ref{lemma62}) into a matrix, the result in (\ref{lemma6}) is obtained. $\hfill\blacksquare$
\end{lemma}
	
\section{Proof of Theorem \ref{theorem_estimate}}
\label{Proof_estimation}
With the given receiving signal ${\bf y}_k^p$, the estimated channel ${\bf h}_k$ based on the LMMSE method is given by
\begin{equation}
	\label{LMMSE}
	\begin{split}
		{\bf {\hat h}}_k &= \mathbb{E}\left\lbrace {\bf {h}}_k \right\rbrace + {\rm {Cov}}\left\lbrace {\bf {h}}_k, {\bf y}_k^p\right\rbrace {\rm {Cov}}^{-1}\left\lbrace {\bf y}_k^p, {\bf y}_k^p\right\rbrace\\
		 & \quad \times ({\bf y}_k^p - \mathbb{E}\left\lbrace {\bf y}_k^p \right\rbrace), 
	\end{split}
\end{equation}
where $ \mathbb{E}\left\lbrace {\bf {h}}_k \right\rbrace$ and $\mathbb{E}\left\lbrace {\bf y}_k^p \right\rbrace$ are equal to 0. By using Lemma 5, we obtain the expressions of ${\text {Cov}}\left\lbrace {\bf {h}}_k, {\bf y}_k^p\right\rbrace$ and ${\text {Cov}}\left\lbrace {\bf y}_k^p, {\bf y}_k^p\right\rbrace$, which is given by
\begin{equation}
	\label{Covhy}
	\begin{split}
	{\mathop{\rm Cov}} \left\{ {{{\bf{h}}_k},{\bf{y}}_k^p} \right\} &= \mathbb{E}\left\{ {\left( {{{\bf{h}}_k} - \mathbb{E}\left\{ {{{\bf{h}}_k}} \right\}} \right){{\left( {{\bf{y}}_k^p - \mathbb{E}\left\{ {{\bf{y}}_k^p} \right\}} \right)}^H}} \right\} \\
%	& = \mathbb{E}\left\{ {{{\left( {{{\bf{C}}^{\rm{B}}}} \right)}^{\frac{1}{2}}}{\bf{\tilde G}}{{\left( {{{\bf{C}}^{\rm{R}}}} \right)}^{\frac{1}{2}}}{\bf{\Phi }}{{\left( {{\bf{C}}_k^{{\rm{RU}}}} \right)}^{\frac{1}{2}}}{{{\bf{\tilde v}}}_k}{\bf{\tilde v}}_k^H{{\left( {{\bf{C}}_k^{{\rm{RU}}}} \right)}^{\frac{1}{2}}}{{\bf{\Phi }}^H}{{\left( {{{\bf{C}}^{\rm{R}}}} \right)}^{\frac{1}{2}}}{{{\bf{\tilde G}}}^H}{{\left( {{{\bf{C}}^{\rm{B}}}} \right)}^{\frac{1}{2}}}} \right\} \\
%	& = \beta _k^{{\rm{RU}}}{\left( {{{\bf{C}}^{\rm{B}}}} \right)^{\frac{1}{2}}}\mathbb{E}\left\{ {{\bf{\tilde G}}{{\left( {{{\bf{C}}^{\rm{R}}}} \right)}^{\frac{1}{2}}}{\bf{\Phi C}}_k^{{\rm{RU}}}{{\bf{\Phi }}^H}{{\left( {{{\bf{C}}^{\rm{R}}}} \right)}^{\frac{1}{2}}}{{{\bf{\tilde G}}}^H}} \right\}{\left( {{{\bf{C}}^{\rm{B}}}} \right)^{\frac{1}{2}}} \\
	& = \beta _k^{{\rm{RU}}}\beta^{{\rm{BR}}}{\rm{Tr}}\left\{ {{\bf{\Phi C}}_k^{{\rm{RU}}}{{\bf{\Phi }}^H}{{\bf{C}}^{\rm{R}}}} \right\}{{\bf{C}}^{\rm{B}}} \\
	&\triangleq \beta _k^{{\rm{RU}}}\beta^{{\rm{BR}}} {\rm{Tr}}\left\{ {\bf Z}_k\right\} {{\bf{C}}^{\rm{B}}},
	\end{split}
\end{equation}
and 
\begin{equation}
	\label{Covyy}
	\begin{split}
		&{\mathop{\rm Cov}} \left\{ {{{\bf{y}}_k},{\bf{y}}_k^p} \right\} \\
		= & \mathbb{E}\left\{ {{{\bf{h}}_k}{\bf{h}}_k^H + \frac{1}{{\tau p^p}}{{\bf{N}}^p}{{\bf{q}}_k}{\bf{q}}_k^H{{\left( {{{\bf{N}}^p}} \right)}^H}} \right\} \\
		= & \beta _k^{{\rm{RU}}}\beta^{{\rm{BR}}} {\rm{Tr}}\left\{ {\bf Z}_k\right\} {{\bf{C}}^{\rm{B}}} + \frac{\sigma^2}{{\tau P_p}}{{\bf{I}}_M}.
	\end{split}
\end{equation}

By substituting the results of (\ref{Covhy}) and (\ref{Covyy}) into (\ref{LMMSE}), we obtain the result in (\ref{hat_hk}).

\section{Proof of Theorem \ref{SINR_T}}
\label{Proof_LB}
The power of desired signal, leakage signal, and the interference are derived, respectively.

The first term ${{{\left| \mathbb{E} {\left\{ {{\bf{h}}_k^H{{{\bf{\hat h}}}_k}} \right\}} \right|}^2}}$ can be given by
\begin{equation}
	\label{first_term}
	\begin{split}
		&{{{\left| \mathbb{E} {\left\{ {{\bf{h}}_k^H{{{\bf{\hat h}}}_k}} \right\}} \right|}^2}} \\ 
		= & {\left| \mathbb{E} {\left\{ {{\bf{h}}_k^H{{\bf{R}}_k}{{\bf{h}}_k}} \right\}} \right|^2} = {\left| {{\rm{Tr}} \left\{ \mathbb{E}{\left\{ {{{\bf{R}}_k}{{\bf{h}}_k}{\bf{h}}_k^H} \right\}} \right\}} \right|^2}\\
		 = &{\left| {{\rm{Tr}}\left\{ {{{\bf{A}}_k}} \right\}} \right|^2},
	\end{split}
\end{equation}
where ${\bf{A}}_k$ is equal to $\beta _k^{{\rm{RU}}}\beta^{{\rm{BR}}} {\rm{Tr}}\left\{ {\bf Z}_k\right\} {{\bf{R}}_k}{{{\bf{C}}^{\rm{B}}}}$.

The second term $\mathbb{E}\left\lbrace {{\left| { {{\bf{h}}_k^H{{{\bf{\hat h}}}_k}  } } \right|}^2}\right\rbrace$ can be calculated as
\begin{equation}
	\label{second_term}
	\begin{split}	
	&\mathbb{E}\left\lbrace {{\left| { {{\bf{h}}_k^H{{{\bf{\hat h}}}_k}  } } \right|}^2}\right\rbrace \\
	= &\mathbb{E}\left\{ {{{\left| {{\bf{h}}_k^H{{\bf{R}}_k}\left( {{{\bf{h}}_k} + \frac{1}{{\sqrt {\tau p^p} }}{{\bf{N}}^p}{{\bf{q}}_k}} \right)} \right|}^2}} \right\} \\
	 = &\mathbb{E}\left\{ {{{\left| {{\bf{h}}_k^H{{\bf{R}}_k}{{\bf{h}}_k}} \right|}^2} + \frac{1}{{\tau P_p}}{{\left| {{\bf{h}}_k^H{{\bf{R}}_k}{{\bf{N}}^p}{{\bf{q}}_k}} \right|}^2}} \right\}.
\end{split}
\end{equation}
By using Lemma 2 in \cite{bjornson2015massive} and Lemma 5, the first term is given by 
\begin{equation}
	\label{first_second}
	\begin{split}
		&\mathbb{E}\left\{ {{{\left| {{\bf{h}}_k^H{{\bf{R}}_k}{{\bf{h}}_k}} \right|}^2}} \right\} \\
		& = {\left( {{\beta ^{\rm BR}}{\beta_k^{\rm RU}}} \right)^2}\mathbb{E}\left\{ {{{\left| {{\bf{\tilde v}}_k^H{{\left( {{\bf{C}}_k^{{\rm{RU}}}} \right)}^{\frac{1}{2}}}{{\bf{\Phi }}^H}{{\left( {{{\bf{C}}^{\rm{R}}}} \right)}^{\frac{1}{2}}}{{{\bf{\tilde G}}}^H}{{\left( {{{\bf{C}}^{\rm{B}}}} \right)}^{\frac{1}{2}}}{{\bf{R}}_k}{{\left( {{{\bf{C}}^{\rm{B}}}} \right)}^{\frac{1}{2}}}{\bf{\tilde G}}{{\left( {{{\bf{C}}^{\rm{R}}}} \right)}^{\frac{1}{2}}}{\bf{\Phi }}{{\left( {{\bf{C}}_k^{{\rm{RU}}}} \right)}^{\frac{1}{2}}}{{{\bf{\tilde v}}}_k}} \right|}^2}} \right\} \\
		& = {\left( {{\beta ^{\rm BR}}{\beta_k^{\rm RU}}} \right)^2}({\left| {{\rm{Tr}}\left\{ {{{\bf{C}}^{\rm{B}}}{{\bf{R}}_k}} \right\}} \right|^2}+{\rm{Tr}}\left\{ {{{\bf{C}}^{\rm{B}}}{\bf{R}}_k^H{{\bf{C}}^{\rm{B}}}{{\bf{R}}_k}} \right\})\mathbb{E} \left\{ |{\bf{\tilde v}}_k^H{\left( {{\bf{C}}_k^{{\rm{RU}}}} \right)^{\frac{1}{2}}}{{\bf{\Phi }}^H}{{\bf{C}}^{\rm{R}}}{\bf{\Phi }}{\left( {{\bf{C}}_k^{{\rm{RU}}}} \right)^{\frac{1}{2}}}{{{\bf{\tilde v}}}_k}|^2\right\}\\
		& = {\left( {{\beta ^{\rm BR}}{\beta_k^{\rm RU}}} \right)^2}({{{\left| {{\rm{Tr}}\left\{ {\bf Z}_k \right\}} \right|}^2} + {\rm{Tr}}\left\{ {\bf Z}_k{\bf Z}_k \right\}} )({{{\left| {{\rm{Tr}}\left\{ {{{\bf{C}}^{\rm{B}}}{{\bf{R}}_k}} \right\}} \right|}^2} + {\rm{Tr}}\left\{ {\bf Z}_k \right\}} ),
	\end{split}
\end{equation}
and the second term in (\ref{second_term}) is
\begin{equation}
	\label{second_second}
	\begin{split}
		&\mathbb{E}\left\{ {{{\left| {{\bf{h}}_k^H{{\bf{R}}_k}{{\bf{N}}^p}{{\bf{q}}_k}} \right|}^2}} \right\} \\
		= &\mathbb{E}\left\{ {{\bf{h}}_k^H{{\bf{R}}_k}{{\bf{N}}^p}{{\bf{q}}_k}{\bf{q}}_k^H{{\left( {{{\bf{N}}^p}} \right)}^H}{{\bf{R}}_k}{{\bf{h}}_k}} \right\} \\
	 = 	&{\rm{Tr}}\left\{\mathbb{E} {\left\{ {{{\bf{h}}_k}{\bf{h}}_k^H} \right\}{{\bf{R}}_k}{\bf{R}}_k^H} \right\} \\
		= &\sigma^2 \beta _k^{{\rm{RU}}}\beta^{{\rm{BR}}}{\rm{Tr}}\left\{ {\bf Z}_k \right\}{\rm{Tr}}\left\{ {{{\bf{C}}^{\rm{B}}}{{\bf{R}}_k}{\bf{R}}_k^H} \right\}. 
	\end{split}
\end{equation}

Then, the interfering term $\mathbb{E}\left\lbrace {{\left| { {{\bf{h}}_k^H{{{\bf{\hat h}}}_{k'}}  } } \right|}^2}\right\rbrace$ is given by 

\begin{equation}
	\label{third}
	\begin{split}
		\mathbb{E}\left\lbrace {{\left| { {{\bf{h}}_k^H{{{\bf{\hat h}}}_{k'}}  } } \right|}^2}\right\rbrace & = \mathbb{E}\left\{ {{\bf{h}}_k^H\mathbb{E}\left\{ {{{{\bf{\hat h}}}_{k'}}{\bf{\hat h}}_{k'}^H} \right\}{{\bf{h}}_k}} \right\} \\
		& = \beta _{k'}^{\rm RU}\beta _k^{\rm RU}{\left( {{\beta ^{\rm BR}}} \right)^2} \times \\
		& \quad \mathbb{E}\left\{\left|{\bf{\tilde v}}_k^H{\left( {{\bf{C}}_k^{{\rm{RU}}}} \right)^{\frac{1}{2}}}{{\bf{\Phi }}^H}{\left( {{{\bf{C}}^{\rm{R}}}} \right)^{\frac{1}{2}}}{{{\bf{\tilde G}}}^H}{\left( {{{\bf{C}}^{\rm{B}}}} \right)^{\frac{1}{2}}}{{\bf{R}}_{k'}}{\left( {{{\bf{C}}^{\rm{B}}}} \right)^{\frac{1}{2}}}{\bf{\tilde G}}{\left( {{{\bf{C}}^{\rm{R}}}} \right)^{\frac{1}{2}}}{\bf{\Phi }}{\left( {{\bf{C}}_{k'}^{{\rm{RU}}}} \right)^{\frac{1}{2}}}{{{\bf{\tilde v}}}_{k'}}\right|^2  \right\} \\
		& \quad + \frac{{\beta ^{\rm BR}}\beta _k^{\rm RU}}{\tau P_p}\mathbb{E}\left\{\left|{\bf{\tilde v}}_k^H{\left( {{\bf{C}}_k^{{\rm{RU}}}} \right)^{\frac{1}{2}}}{{\bf{\Phi }}^H}{\left( {{{\bf{C}}^{\rm{R}}}} \right)^{\frac{1}{2}}}{{{\bf{\tilde G}}}^H}{\left( {{{\bf{C}}^{\rm{B}}}} \right)^{\frac{1}{2}}}{{\bf{R}}_{k'}}{{\bf{N}}^p}{{\bf{q}}_{k'}}\right|^2  \right\}.
	\end{split}	
\end{equation}

The first term in (\ref{third}) is given by
\begin{equation}
	\label{first_third}
	\begin{split}
		 &\mathbb{E}\left\{\left|{\bf{\tilde v}}_k^H{\left( {{\bf{C}}_k^{{\rm{RU}}}} \right)^{\frac{1}{2}}}{{\bf{\Phi }}^H}{\left( {{{\bf{C}}^{\rm{R}}}} \right)^{\frac{1}{2}}}{{{\bf{\tilde G}}}^H}{\left( {{{\bf{C}}^{\rm{B}}}} \right)^{\frac{1}{2}}}{{\bf{R}}_{k'}}{\left( {{{\bf{C}}^{\rm{B}}}} \right)^{\frac{1}{2}}}{\bf{\tilde G}}{\left( {{{\bf{C}}^{\rm{R}}}} \right)^{\frac{1}{2}}}{\bf{\Phi }}{\left( {{\bf{C}}_{k'}^{{\rm{RU}}}} \right)^{\frac{1}{2}}}{{{\bf{\tilde v}}}_{k'}}\right|^2  \right\} \\
		 & = {\left| {{\rm{Tr}}\left\{ {{{\bf{R}}_{k'}}{{ {{{\bf{C}}^{\rm{B}}}} }}} \right\}} \right|^2}\mathbb{E}\left\{ {{\bf{\tilde v}}_k^H{{\left( {{\bf{C}}_k^{{\rm{RU}}}} \right)}^{\frac{1}{2}}}{{\bf{\Phi }}^H}{{ {{{\bf{C}}^{\rm{R}}}} }} {\bf{\Phi C}}_{k'}^{{\rm{RU}}}{{\bf{\Phi }}^H}}{{ {{{\bf{C}}^{\rm{R}}}} }}{{\bf{\Phi }}{{\left( {{\bf{C}}_k^{{\rm{RU}}}} \right)}^{\frac{1}{2}}}{{{\bf{\tilde v}}}_k}}\right\} \\
		 & \quad + {\rm{Tr}}\left\{ {\bf Z}_{k'} \right\}{\rm{Tr}}\left\{ {{{\bf{C}}^{\rm{B}}}{{\bf{R}}_{k'}}{{\bf{C}}^{\rm{B}}}{\bf{R}}_{k'}^H} \right\} \mathbb{E}\left\{ {{\bf{\tilde v}}_k^H{{\left( {{\bf{C}}_k^{{\rm{RU}}}} \right)}^{\frac{1}{2}}}{{\bf{\Phi }}^H}{{ {{{\bf{C}}^{\rm{R}}}} }}}{{\bf{\Phi }}{{\left( {{\bf{C}}_k^{{\rm{RU}}}} \right)}^{\frac{1}{2}}}{{{\bf{\tilde v}}}_k}}\right\} \\
		 & = {{{\left| {{\rm{Tr}}\left\{ {{{\bf{C}}^{\rm{B}}}{{\bf{R}}_{k'}}} \right\}} \right|}^2}{\rm{Tr}}\left\{ {\bf Z}_k {\bf Z}_{k'}\right\} + {\rm{Tr}}\left\{ {\bf Z}_{k'} \right\}{\rm{Tr}}\left\{ {{{\bf{C}}^{\rm{B}}}{{\bf{R}}_{k'}}{{\bf{C}}^{\rm{B}}}{\bf{R}}_{k'}^H} \right\}{\rm{Tr}}\left\{ {\bf Z}_k \right\}},
	\end{split}
\end{equation}

and the second term in (\ref{third}) is
\begin{equation}
	\label{second_third}
	\begin{split}
		&\mathbb{E}\left\{\left|{\bf{\tilde v}}_k^H{\left( {{\bf{C}}_k^{{\rm{RU}}}} \right)^{\frac{1}{2}}}{{\bf{\Phi }}^H}{\left( {{{\bf{C}}^{\rm{R}}}} \right)^{\frac{1}{2}}}{{{\bf{\tilde G}}}^H}{\left( {{{\bf{C}}^{\rm{B}}}} \right)^{\frac{1}{2}}}{{\bf{R}}_{k'}}{{\bf{N}}^p}{{\bf{q}}_{k'}}\right|^2  \right\}\\
		& = \sigma^2{\rm{Tr}}\left\{ {{{\bf{C}}^{\rm{B}}}{{\bf{R}}_{k'}}{\bf{R}}_{k'}^H} \right\} \mathbb{E}\left\{{\bf{\tilde v}}_k^H{\left( {{\bf{C}}_k^{{\rm{RU}}}} \right)^{\frac{1}{2}}}{{\bf{\Phi }}^H}{{\bf{C}}^{\rm{R}}}{\bf{\Phi }}{\left( {{\bf{C}}_k^{{\rm{RU}}}} \right)^{\frac{1}{2}}}{{{\bf{\tilde v}}}_k}\right\} \\
		& = \sigma^2{\rm{Tr}}\left\{ {{{\bf{C}}^{\rm{B}}}{{\bf{R}}_{k'}}{\bf{R}}_{k'}^H} \right\}{\rm{Tr}}\left\{ {\bf Z}_k \right\}.
	\end{split}
\end{equation}

By substituting $\mathbb{E}\left\lbrace \left|n_k\right|^2\right\rbrace = \sigma^2$ and these expressions into (\ref{SINR}), we obtain the expression of SINR.

\section{Proof of Lemma \ref{lemma2_gradient}}
\label{Proof_lemma2}
With the given matrices ${\bf A}$ and ${\bf B}$, we substitute ${\bf \Phi} = {\rm diag}({\bf b})$ into ${\rm{Tr}}\left\{ {{\bf{A}}{\bf \Phi} {\bf{B}}{{\bf \Phi} ^H}} \right\}$ and have a more explicit form, which is given by
\begin{equation}
	\label{rewriting_lemma2}
	\begin{split}
		{\rm{Tr}}\left\{ {{\bf{A}}{\bf \Phi} {\bf{B}}{{\bf \Phi} ^H}} \right\}& = \sum\limits_i {\sum\limits_j {{{\left[ {\bf{A}} \right]}_{ij}}{{\left[ {\bf \Phi}  \right]}_{jj}}{{\left[ {\bf{B}} \right]}_{ji}}{{\left[ {{{\bf \Phi}^H}} \right]}_{ii}}} } \\
		& = {{\bf{b}}^H}\left( {{\bf{A}} \odot {\bf{B}}^{T}} \right){\bf{b}}.
	\end{split}
\end{equation}

Then, based on the result in (\ref{rewriting_lemma2}), it is convenient to derive the gradient in terms of the $n$th element $\theta_n$ by using the chain rule, which is given by
\begin{equation}
	\label{n_gradient}
	\begin{split}
		\frac{{\rm{Tr}}\left\{ {{\bf{A}}{\bf \Phi} {\bf{B}}{{\bf \Phi} ^H}} \right\}}{{\partial \theta_n }} & = - j{e^{ - j{\theta _n}}}{\left[ {\left( {{\bf{A}} \odot {{\bf{B}}^T}} \right){\bf{b}}} \right]_n} \\
		&\quad + j{\left[ {{{\bf{b}}^H}\left( {{\bf{A}} \odot {{\bf{B}}^T}} \right)} \right]_n}{e^{j{\theta _n}}}.
	\end{split}
\end{equation}

According to the above result of the $n$th element's gradient, the proof is completed by collecting (\ref{n_gradient}) into a vector.

\section{Proof of Lemma \ref{lemma3_gradient}}
\label{Proof_lemma3}
For ease of expression, we define ${{\bf{C}}^{\rm{B}}}{\left( {\beta _k^{{\rm{RU}}}\beta^{{\rm{BR}}}{\rm{Tr}}\left\{ {\bf Z}_k \right\}{{\bf{C}}^{\rm{B}}} + \frac{\sigma^2}{{\tau P_p}}{{\bf{I}}_M}} \right)^{ - 1}}$ as ${\bf W}_k$. Then, we obtain the gradient of ${\rm{Tr}}\left\{ {\bf{X}}{\bf R}_k \right\}$ with respect to the $n$th element $\theta_n$ by using the chain rule, which is detailed as
\begin{equation}
	\label{lemma3_n_gradient}
	\begin{split}
	\frac{{\rm{Tr}}\left\{ {\bf{X}}{\bf R}_k \right\}}{{\partial \theta_n }} & = \frac{ \partial \left( {\beta _k^{{\rm{RU}}}\beta ^{{\rm{BR}}}{\rm{Tr}}\left\{ {\bf Z}_k \right\} } \right)  {\rm{Tr}} \left\lbrace  {\bf X}{\bf W}_k\right\rbrace }{\partial \theta_n}\\
	& = \beta _k^{{\rm{RU}}}\beta ^{{\rm{BR}}} \frac{\partial {\rm{Tr}}\left\{ {\bf Z}_k \right\}}{\partial \theta_n} {\rm{Tr}} \left\lbrace  {\bf X}{\bf W}_k \right\rbrace \\
	&\quad +  {\beta _k^{{\rm{RU}}}\beta ^{{\rm{BR}}}{\rm{Tr}}\left\{ {\bf Z}_k \right\} }\frac{\partial {\rm{Tr}} \left\lbrace  {\bf X}{\bf W}_k \right\rbrace}{\partial \theta_n}. \\
	\end{split}
\end{equation}
The first term in (\ref{lemma3_n_gradient}) can be readily obtained based on the result in (\ref{n_gradient}). Then, the second term can be calculated by using $\partial ({\bf A})^{-1} = -({\bf A})^{-1}\partial ({\bf A})({\bf A})^{-1}$, which is given by
\begin{equation}
	\label{second_lemma3}
	\begin{split}
		&\frac{\partial {\rm{Tr}} \left\lbrace  {\bf X}{{\bf{C}}^{\rm{B}}}{\left( {\beta _k^{{\rm{RU}}}\beta^{{\rm{BR}}}{\rm{Tr}}\left\{ {\bf Z}_k \right\}{{\bf{C}}^{\rm{B}}} + \frac{\sigma^2}{{\tau p^p}}{{\bf{I}}_M}} \right)^{ - 1}} \right\rbrace}{\partial \theta_n} \\
		= &-{{\rm{Tr}} \left\lbrace  {\bf X}{\bf W}_k {\bf W}_k\right\rbrace} \times \beta _k^{{\rm{RU}}}\beta ^{{\rm{BR}}}\frac{\partial {\rm{Tr}}\left\{ {\bf Z}_k \right\}}{\partial \theta_n}.
	\end{split}
\end{equation}

Substituting the results of (\ref{second_lemma3}) into (\ref{lemma3_n_gradient}) and combining the $n$th element's gradient to a vector, we have
\begin{equation}
	\label{final_lemma3}
	\begin{split}
		\frac{{\rm{Tr}}\left\{ {\bf{X}}{\bf R}_k \right\}}{{\partial \boldsymbol{\theta}  }} &= \beta _k^{{\rm{RU}}}\beta ^{{\rm{BR}}} \times {\bf u}_g({\bf C}^{\rm R},{\bf C}^{\rm RU}_k) \times \Big( {\rm{Tr}} \left\lbrace  {\bf X}{\bf W}_k  \right\rbrace \\
		&\quad - \beta _k^{{\rm{RU}}}\beta ^{{\rm{BR}}}{\rm{Tr}}\left\{ {\bf Z}_k \right\} {\rm{Tr}} \left\lbrace  {\bf X}{\bf W}_k {\bf W}_k  \right\rbrace \Big) .
	\end{split}
\end{equation} 
The result in Lemma \ref{lemma3_gradient} is completed.

\end{appendices}	
\bibliographystyle{IEEEtran}
\bibliography{RISrelated}

\end{document}